\documentclass[12pt, draftclsnofoot, onecolumn]{IEEEtran}
\usepackage{cite}
\usepackage{graphicx}

\DeclareGraphicsExtensions{.pdf}

\usepackage{algpseudocode,algorithmicx}
\usepackage[linesnumbered,ruled]{algorithm2e}
\let\oldnl\nl
\newcommand{\nonl}{\renewcommand{\nl}{\let\nl\oldnl}}
\usepackage{amsmath}
\usepackage{amsfonts}
\usepackage{array}
\usepackage{url}
\usepackage{cases}
\usepackage{booktabs}
\usepackage{multirow}
\usepackage{siunitx}
\usepackage[printonlyused,withpage]{acronym}

\acrodef{ms}{millisecond}
\acrodef{ge2e}{Generalized end-to-end}
\acrodef{IoT}{internet of things}
\acrodef{IIoT}{industrial internet of things}
\acrodef{GDPR}{general data protection regulation}
\acrodef{MFB}{mel-filterbank}
\acrodef{VAD}{voice activity detection}
\acrodef{LSTM}{long short-term memory}
\acrodef{MLP}{multilayer perceptron}
\acrodef{ReLU}{rectified linear unit}
\acrodef{SGD}{stochastic gradient descent}
\acrodef{Adam}{adaptive moment estimation}
\acrodef{t-SNE}{t-distributed stochastic neighbor embedding}
\acrodef{EER}{equal error rate}
\acrodef{minDCF}{minimum of the normalized detection cost function}

\begin{document}

\title{Dynamic Recognition of Speakers for Consent Management by Contrastive Embedding Replay}

\author{\IEEEauthorblockN{Arash~Shahmansoori\IEEEauthorrefmark{1}\IEEEauthorrefmark{2} and
Utz~Roedig\IEEEauthorrefmark{1},~\IEEEmembership{Member,~IEEE}}
\IEEEauthorblockA{
\thanks{\IEEEauthorrefmark{1}A. Shahmansoori and U. Roedig are with the School of Computer Science and Information Technology, Western Gateway Building, University College Cork, Cork,
Ireland, emails: arash.mansoori65@gmail.com and u.roedig@cs.ucc.ie.}
\thanks{\IEEEauthorrefmark{2} Corresponding author: Arash Shahmansoori}
\thanks{Note: This work has been submitted to the IEEE for possible publication. Copyright may be transferred without notice, after which this version may no longer be accessible.}
}
}

\maketitle

\begin{abstract}

Voice assistants overhear conversations and a consent management mechanism is required. Consent management can be implemented using speaker recognition. Users that do not give consent enrol their voice and all their further recordings are discarded. Building speaker recognition-based consent management is challenging as dynamic registration, removal, and re-registration of speakers must be efficiently handled. This work proposes a consent management system addressing the aforementioned challenges. A contrastive based training is applied to learn the underlying speaker equivariance inductive bias. The contrastive features for buckets of speakers are trained a few steps into each iteration and act as replay buffers. These features are progressively selected using a multi-strided random sampler for classification. Moreover, new methods for dynamic registration using a portion of old utterances, removal, and re-registration of speakers are proposed. The results verify memory efficiency and dynamic capabilities of the proposed methods and outperform the existing approach from the literature.

\end{abstract}

\begin{IEEEkeywords}
Consent management, voice assistant systems, contrastive embedding replay, multi-strided sampling, dynamic learning.
\end{IEEEkeywords}

\IEEEpeerreviewmaketitle

\section{Introduction} \label{intro}
\IEEEPARstart{M}{any} recent \ac{IoT} applications such as smart homes, smart transport systems or smart healthcare rely on voice assistants as primary user interface. This is due to the fact that end-users prefer to communicate with \ac{IoT} devices more naturally, using voice commands rather than classical interfaces such as a touch screen \cite{8246828}. Consequently, consent management is now becoming a concern. For example, the recent European Union legislation, \ac{GDPR}, requires all parties' consent for personal data collection. In the context of voice assistant systems, providing this feature is essential to protect users from being recorded without giving consent. If not giving consent, users should at least be able to communicate dissent such that their voice is not recorded. Implementing such a consent/dissent management system for voice assistants is challenging. The existence of voice assistant systems to nearby users may initially not be evident. Also, there is no obvious interface to articulate consent or dissent. Recent initial ideas to implement consent management can be divided in two broad categories:

\begin{enumerate}
\item Consent management without voice assistant support
\item Consent management with voice assistant support
\end{enumerate}

The first category assumes that the operators of a voice assistant ecosystem do not support the implementation of consent management while the second approach assumes collaboration of a provider, e.g., Amazon in the case of the Echo voice assistant. In the first category, Denial of Service approaches have been proposed; the voice assistant is prevented to collect voice samples by a non-consenting party. Specifically, an acoustic jamming device can be used to prevent all voice assistant systems in the vicinity of a user to record \cite{cheng2018towards}. While such approach is possible, it is difficult to implement reliably in practical settings. In the second category more options are available. One approach is to add information to the acoustic channel that can subsequently be detected by a back-end. A sound signal, i.e., a tag, is embedded in the audio stream via a speaker that can be used by the voice assistant's back-end for consent management \cite{Cheng_tagging}. This approach faces challenges, in particular when consent of multiple users should be handled, requiring collision management of tag signals. A second approach in this category is the use of speaker recognition for consent management. However, the direct use of such approaches in the context of consent management is not practical as will be briefly discussed.

In \cite{Snell2017PrototypicalNF,Vinyals2016MatchingNF,Finn2017ModelAgnosticMF}, few-shot learning methods are used to generalize on the classes with similar features never seen during the training mode for speaker recognition. However, in the context of consent management for voice assistant systems such a generalization actually hurts the consent management as a privacy measure. This is due to the fact that there is a possibility for generalizing to speakers that are already providing their consent according to the samples from the speakers that do not. In \cite{ven_brain-inspired_2020,Oswald2020Continual,DBLP:conf/cvpr/MaiLKS21}, replay based buffer methods for continually learning a set of tasks are proposed such that each time the network only has full access to the data for the current task. However, these approaches usually require difficult ways to generate the replay, learn the parameters of a target network, and sampling the buffer in the input space leading to slow convergence, performance degradation, computationally complex operations, and large memory requirements. Moreover, it is assumed that the entire data for each task is provided sequentially and the network is fully trained for the current task using the replay buffer of the previous tasks to avoid catastrophic forgetting. This is not necessarily the case for the consent management systems as generally only a small portion of dataset for each bucket of speakers may be provided during each iteration.

In \cite{ICASSP_GE2E,Sang2022,Zhang2021ContrastiveSL,9414973,alexeyhousehold2022}, different methods are applied for speaker verification systems. Such applications usually require large batch size, larger models, and full access to the entire utterances of speakers during training. However, this leads to slow convergence, large memory requirements, and performance degradation with partial access to utterances of speakers for consent management in voice assistant systems. Moreover, the main concern for speaker verification systems is the existence of speakers in a pool of previously registered speakers. This can be useful for certain applications that only need to screen a set of speakers and verify their existence \cite{912681}.  

In the context of consent management for voice assistant systems, the main concerns are about the dynamic management of consent for ``the specific speakers'' in ``the specific buckets'', efficient use of their utterances, and not storing their private information unnecessarily in the back-end during new registrations. Moreover, it is totally possible for speakers not to provide their consent for certain attributes, e.g., gender, but providing their consent for other attributes, e.g., transcribing their speech. In other words, ``identifying'' the speakers who do not provide consent is of particular interest as they may provide their consent for certain attributes. In conclusion, it is not a zero-sum game to verify the existence of speakers or screen a given set of speakers, but rather a dynamic process to manage their consent and identify them.

The specific contributions are summarized as follows.
\begin{enumerate}
\item A training process based on the contrastive embeddings as a way to learn speaker equivariance inductive bias is proposed. The proposed approach is efficient in terms of convergence speed and accurate prediction of speakers that do not provide consent. This is mainly due to learning the underlying speaker equivariance inductive biases and using them as replay buffer continuously during the training for classification.
\item A progressive multi-strided random sampling of the contrastive embedding replay buffer is proposed. The proposed sampling strategy starts with the large number of utterances from the initial buckets to fill up the memory size. Then, it sparsely samples the buckets of speakers to preserve enough memory for the buckets seen so far. This leads to memory efficiency, progressive increase of task difficulty, and avoiding parameter shift to the buckets of speakers with more samples.
\item A dynamic algorithm for registering new speakers in different buckets is proposed. The new speakers are registered, using only a portion of the utterances of old speakers, in the unique buckets, obtained according to $L2$ pairwise distance from the prototypes of the previous registrations, in each round. This is achieved using a dynamic programming with linear time complexity.
\item A dynamic algorithm for removing the previously registered speakers from the pool of speakers is proposed. The proposed algorithm is capable of selectively forgetting the previously learned contrastive features for speakers in different buckets with the reduced elapsed time. Also, the proposed method can quickly re-register the removed speakers in case this is required.
\item All the aforementioned points are applied for both supervised and unsupervised modes.
\end{enumerate}

\begin{figure}[t]
    \centering
    \includegraphics[scale=0.5]{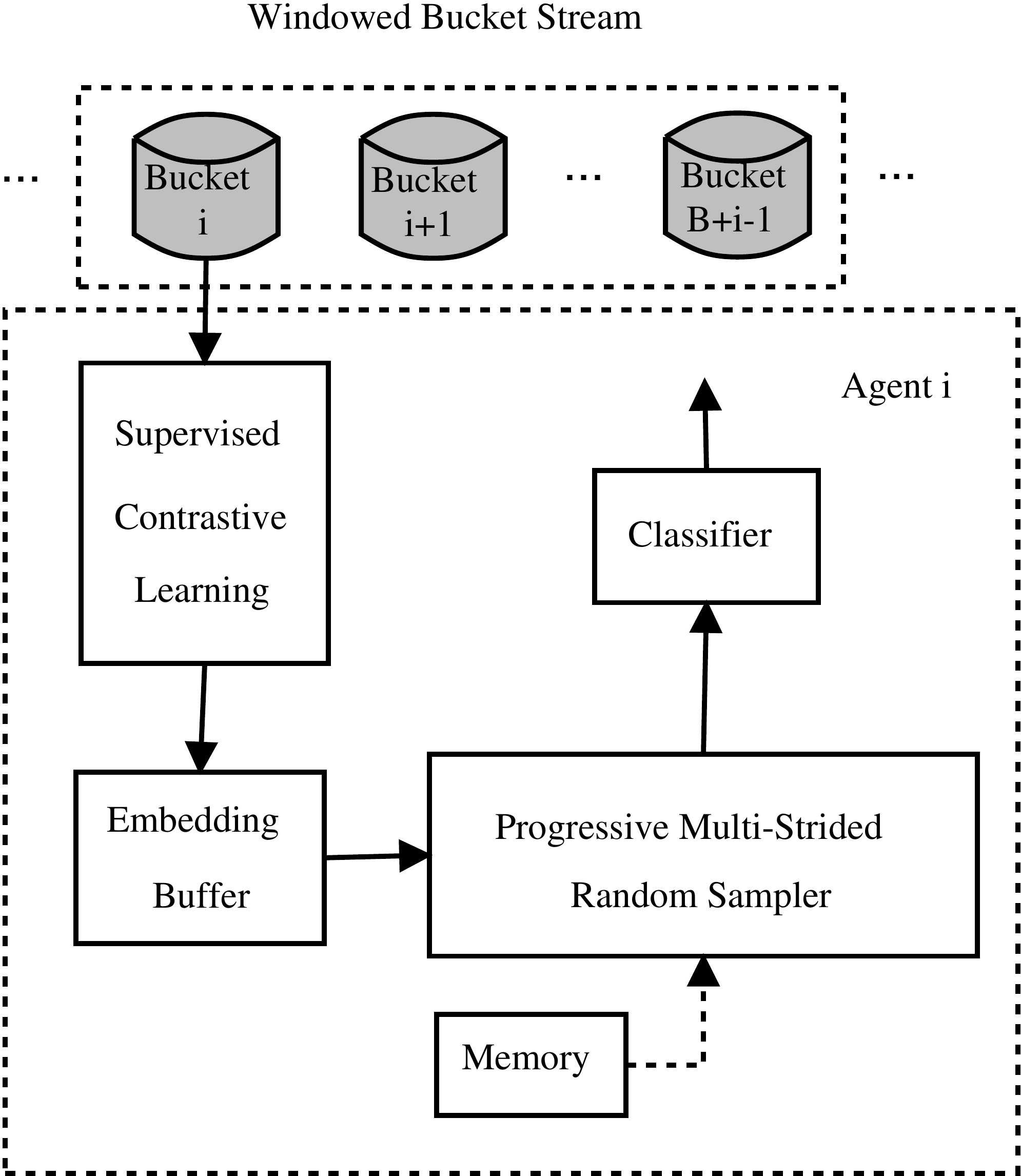}
    \caption{The process for the proposed training with contrastive embedding replay for an agent.}
    \label{ContrativeEmbeddingTraining}
\end{figure}

\section{Method}\label{method}
In this section, first the framework for training of speakers for consent management is explained in an algorithmic way. Then, a mechanism for dynamic registration of new speakers' consent is proposed. Finally, a method for removing the previously registered speakers' consent is developed.

\subsection{Training}

The framework for the entire training process is described by an agent interacting with groups of speakers, i.e., buckets. Each agent is responsible for training a windowed stream of buckets in a modular manner. This way, it is possible to distribute the training process among different agents. Fig. \ref{ContrativeEmbeddingTraining} shows the overall pictorial viewpoint of the proposed training with contrastive embedding replay in the supervised mode. The extension of the proposed approach to the unsupervised mode is provided in the results.

The proposed method starts the training with the selection of buckets of speakers that do not provide consent, i.e., a windowed bucket stream, for each agent. Subsequently, each bucket of speakers is fed to a supervised contrastive learning framework where feature extraction is achieved by running the supervised contrastive learning for each bucket individually only a few steps into the training. Note that contrastive feature extraction in each run of the training process requires only a few steps, e.g., $\mathrm{epochs}_{\mathrm{cont}}=5$. In other words, the proposed method does not wait for the full convergence of the supervised contrastive learning for each bucket during each run of the training process. The individual supervised contrastive loss $\mathcal{L}_{\mathrm{sup}}^{(b)}$ for bucket $b\in\tilde{\mathbf{b}}_{i}$, where $\tilde{\mathbf{b}}_{i}=[i, i+1, \ldots, B+i-1]$ denotes the list of $B$ buckets for the Agent $i$, is defined as follows.
\begin{equation}
\mathcal{L}_{\mathrm{sup}}^{(b)}  =  \sum_{(s,u)\in\mathcal{I}_{b}} \mathcal{L}_{\mathrm{sup},s,u}^{(b)},
\label{sup_cont_loss}
\end{equation}
where $\mathcal{I}_{b}$ denotes all the speakers' utterances in the batch for the bucket $b$ during training, and $\mathcal{L}_{\mathrm{sup},s,u}^{(b)}$ is defined as follows. 
\begin{IEEEeqnarray*}{l}
\mathcal{L}_{\mathrm{sup},s,u}^{(b)} = \frac{-1}{\vert\mathcal{P}_{s,b}(u)\vert}
\sum_{p\in\mathcal{P}_{s,b}(u)}\log\left(\frac{\exp\left(\mathbf{z}^{(u)}_{s,b}\cdot\mathbf{z}_{p}/\tau\right)}{\sum_{a\in\mathcal{A}_{s,b}(u)}\exp\left(\mathbf{z}^{(u)}_{s,b}\cdot\mathbf{z}_{a}/\tau\right)}\right),
\IEEEyesnumber
\label{sup_cont_lossb}
\end{IEEEeqnarray*}
where $\mathcal{A}_{s,b}(u)$ and $\mathcal{P}_{s,b}(u)$ are defined as
\begin{IEEEeqnarray}{rCl}
\mathcal{A}_{s,b}(u) & := & \mathcal{I}_{s,b}\backslash\{u\}\label{set_definitions_A},\\
\mathcal{P}_{s,b}(u) & := & \mathcal{P}_{s,b}\backslash\{u\}\label{set_definitions_P},
\end{IEEEeqnarray}
in which $\mathcal{I}_{s,b}$ and $\mathcal{P}_{s,b}$ denote all the utterances of other speakers $\tilde{s}\neq s$ in the bucket $b$, and all the utterances of speaker $s$ in the bucket $b$, respectively. The operation $\backslash\{u\}$ excludes the anchor utterance $u$ from the corresponding set, and $\vert\mathcal{P}_{s,b}(u)\vert$ denotes the corresponding cardinality of the set of utterances for speaker $s$ in bucket $b$ excluding anchor utterance $u$. The parameter $\tau$ is a positive scalar denoting the temperature. The embedding terms in \eqref{sup_cont_lossb} are obtained as
\begin{equation}\label{enc}
\mathbf{z}_{s,b} = \mathrm{Proj}_{\theta_{\mathrm{proj},b}}(\mathrm{Emb}_{\theta_{\mathrm{e},b}}(\mathbf{x}_{s,b}))=\mathrm{Enc}_{\theta_{b}}(\mathbf{x}_{s,b}),
\end{equation}
where $\mathrm{Emb}_{\theta_{\mathrm{e},b}}(.)$ and $\mathrm{Proj}_{\theta_{\mathrm{proj},b}}(.)$ denote the embedding network and the projection head, respectively. The notation $\mathrm{Enc}_{\theta_{b}}(.)$ is used for the encoder containing the embedding network followed by the projection head. The projection head is implemented using the attention pooling layer to obtain the embeddings for speaker $s$ in bucket $b$, $\mathbf{z}_{s,b}$, with the parameter set $\theta_{b}=\{\theta_{\mathrm{e},b},\theta_{\mathrm{proj},b}\}$. The embedding $\mathbf{z}_{s,b}$ contains the elements $\mathbf{z}^{(j)}_{s,b}$ for the $j$-th utterance, $\mathbf{z}_{p}$/$\mathbf{z}_{a}$ denotes the corresponding positive/negative embedding, and $\mathbf{x}_{s,b}$ denotes the input features obtained as described in the simulations.

\begin{figure}[t]
    \centering
    \includegraphics[scale=0.5]{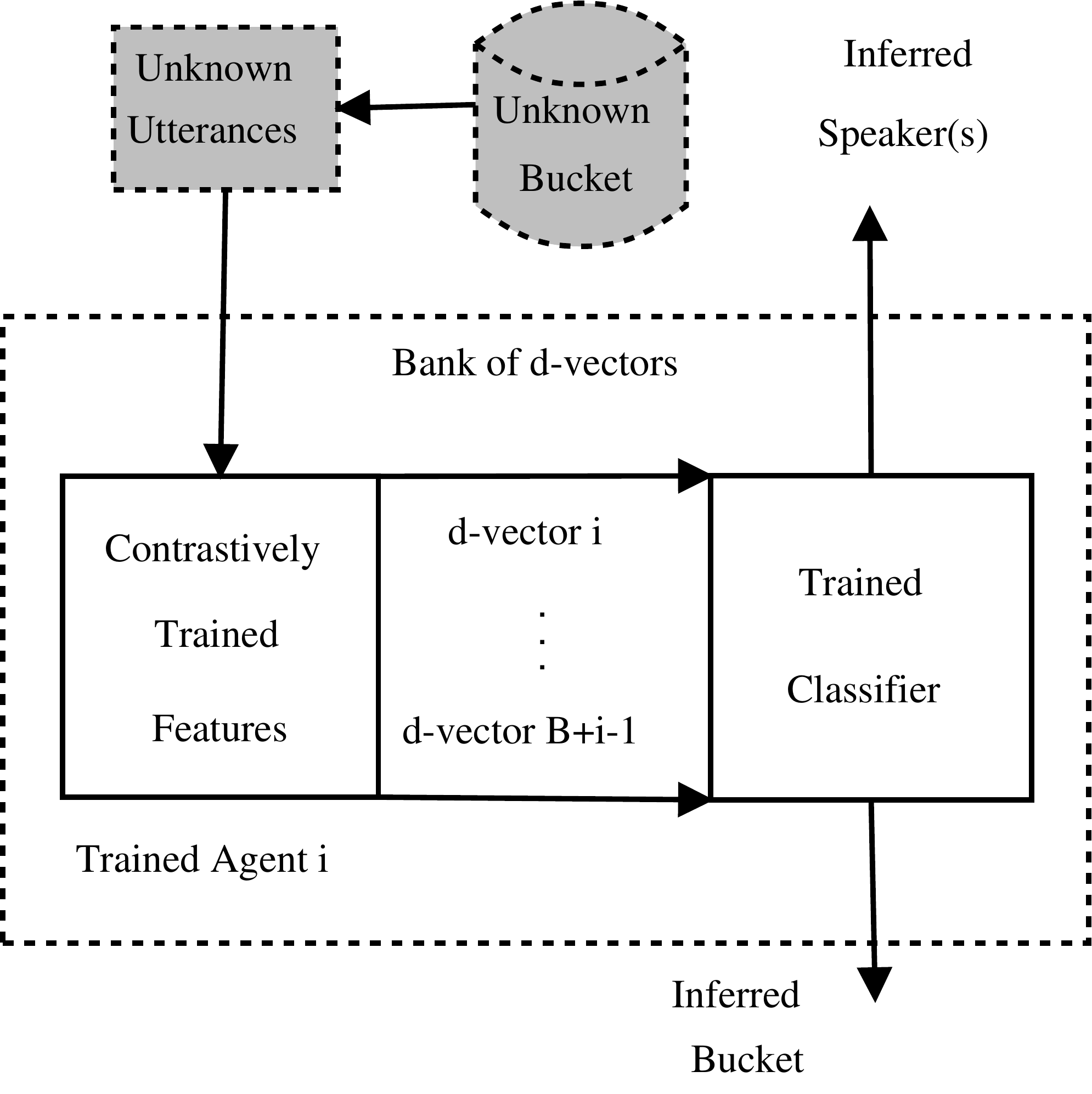}
    \caption{Pictorial viewpoint of the proposed method in the inference mode for a given agent.}
    \label{ContrativeEmbeddingInference}
\end{figure}

Subsequently, the contrastive embedding buffer is sampled according to a progressive multi-strided random sampling algorithm, described by a collection of functions in the Appendix. \ref{app:multi-stride-sampling}. Finally, the classifier is trained using the samples provided by the aforementioned progressive multi-strided random embedding buffer sampling algorithm. In other words, the contrastive training provides an inductive bias for speaker classification during training as the main task. Presenting the contrastive inductive bias to the main classification task during training results efficient use of data and fast convergence as will be discussed in the simulations. Fig. \ref{ContrativeEmbeddingInference} represents the proposed method after training in the inference mode. In this mode, utterances of unknown bucket of unknown speaker(s) are provided to the trained agent. Using the supervised contrastively trained feature extraction, a bank of d-vectors is achieved that can be used as the inputs to the trained classifier for inferring the speaker(s) together with the corresponding bucket. To simplify the notation, the subscript $i$ for the bucket list $\tilde{\mathbf{b}}_{i}$ is dropped for the rest of the manuscript. 

\begin{algorithm}[!t]\label{cnst_emb_rep}

\SetAlgoLined
\DontPrintSemicolon

\SetKwFunction{nutts}{$num_{spk,utts}$}
\SetKwFunction{ncollection}{$collection_{indx}$}
\SetKwFunction{interbkt}{$sample_{int-bkt}$}

\SetKwFunction{emetrics}{$eval_{\mathrm{metric},b}$}

Compute $n_{spk,\mathrm{utt}}$ according to \eqref{num_spk_utts}.

\For {$epoch$ in $range(epochs)$}{
Obtain $\mathcal{C}_{indx}$ according to \eqref{indx_collect}.\;
$\mathbf{zy}_{\mathrm{init}}$ $=$ ([], [])\;
\For {$\_$, $b$ in $enumerate(\tilde{\mathbf{b}})$}{
Load a random shard of dataset for speakers in $b$ with $n_{\mathrm{utt}}$.

\If{!(early-stop$_{b}$)}
{Train $\mathrm{Enc}_{\theta_{\mathrm{b}}}(.)$ for $\mathrm{epochs}_{\mathrm{cont}}$ contrastively and save checkpoints.}

Return the embeddings in \eqref{enc} for latest checkpoints and corresponding labels.

$\mathbf{D}^{max_{\mathrm{mem}}}_{\mathrm{buff}}$, $\mathbf{y}^{max_{\mathrm{mem}}}_{\mathrm{buff}}$ $=$ \interbkt{$\mathcal{C}_{indx}[b]$, $\mathbf{zy}_{b}$, $\mathbf{zy}_{\mathrm{init}}$}

Train $\mathrm{Cls}_{\phi}(.)$ using $\{\mathbf{D}^{max_{\mathrm{mem}}}_{\mathrm{buff}}, \mathbf{y}^{max_{\mathrm{mem}}}_{\mathrm{buff}}\}$ for $\mathrm{epochs}_{\mathrm{cls}}$ and save checkpoints.
}

Progressively evaluate \emetrics{.} for $b\in\tilde{\mathbf{b}}$.

Update ``early-stop$_{b}$'' parameters according to \emetrics{.}.

\If{early-stop$_{\tilde{\mathbf{b}}[-1]}$}{Break the training.}

}
\caption{Consent management with contrastive embedding replay}
\end{algorithm}

The entire process of consent management is proposed in the Algorithm. \ref{cnst_emb_rep}. After initializing the parameters of contrastive feature extraction encoder $\mathrm{Enc}_{\{\theta_{\mathrm{b}}\}}(.)$, $\{\theta_{\mathrm{b}}\}$, and classifier $\mathrm{Cls}_{\phi}(.)$, $\phi$, the list $\mathbf{n}^{\mathrm{reg}}_{bkt}\in\{0, 1\}^{B}$ containing the number of new speakers per buckets, with the values selected from the set $\{0, 1\}$, as will be explained in dynamic registration procedure, is set to zero. This is due to the fact that there are no new registered speakers in the pool of speakers for consent management. In step 1, the number of utterances per speaker $n_{spk,\mathrm{utt}}$ is obtained as follows.
\begin{equation}\label{num_spk_utts}
n_{spk,\mathrm{utt}} = num_{spk,utts}(max_{\mathrm{mem}}, \mathbf{n}_{bkt}, \mathbf{n}^{\mathrm{reg}}_{bkt}), 
\end{equation}
where the function $num_{spk,utts}(.)$, defined in the Appendix. \ref{app:multi-stride-sampling}, computes $n_{spk,\mathrm{utt}}$ according to the maximum allowed memory size $max_{\mathrm{mem}}$ for training. The argument $\mathbf{n}_{bkt}\in\mathbb{Z}^{B}_{\geq 0}$ denotes a list of length $B$ containing a non-negative set of integers representing the number of speakers per buckets. The training iterations over the specified range of $epochs$ starts at step 2. Prior to starting the registration of speakers in the buckets, a dictionary, with the keys of bucket $b$ and values of the flattened list of indices of utterances per speakers per bucket $b$, is obtained in step 3 as follows.
\begin{equation}\label{indx_collect}
\mathcal{C}_{indx} = collection_{indx}(n_{s,\mathrm{utt}}, n_{spk,\mathrm{utt}}, \mathbf{n}_{bkt}, \mathbf{n}^{\mathrm{reg}}_{bkt}),
\end{equation}
where the function $collection_{indx}(.)$ is defined in the Appendix. \ref{app:multi-stride-sampling}, and $n_{s,\mathrm{utt}}$ denotes the number of selected utterances per speaker. A tuple of speaker embeddings and corresponding labels are set to empty lists in step 4. The iterations over $\tilde{\mathbf{b}}$ starts in step 5 where the $enumerate(.)$ generates a specific bucket $b$ for each iteration. In step 6, a random shard of dataset for $n_{\mathrm{utt}}$ utterances per speakers in $b$ for each $epoch$ is loaded. In case the early stopping status obtained according to the progressive evaluation of a given metric, e.g., accuracy or loss, up to bucket $b\in\tilde{\mathbf{b}}$ is not true, denoted as !(early-stop$_{b}$) where !(.) negates the logical statement in parenthesis, train $\mathrm{Enc}_{\theta_{\mathrm{b}}}(.)$ for $\mathrm{epochs}_{\mathrm{cont}}$ steps contrastively, steps 7--9. This is due to the fact that the task difficulty is progressively increased during each iteration. In particular, the number of speakers is increased by providing samples from each bucket progressively, and the number of utterances per speaker is decreased as a result of the $max_{\mathrm{mem}}$ memory budget. In other words, if the classifier is able to distinguish different classes with sufficiently high accuracy for harder tasks, according to the corresponding contrastively trained features, it is also able to classify the simpler tasks prior to that task.

Next, the speaker embeddings are obtained according to \eqref{enc} for the latest available checkpoints and the corresponding labels are returned for $b$ in step 10. Using the inter-bucket sampling function $sample_{int-bkt}(.)$, described in the Appendix. \ref{app:multi-stride-sampling}, the progressive features $\mathbf{D}^{max_{\mathrm{mem}}}_{\mathrm{buff}}$ and corresponding labels $\mathbf{y}^{max_{\mathrm{mem}}}_{\mathrm{buff}}$ with $max_{\mathrm{mem}}$ memory size are obtained in step 11. The progressive features and the corresponding labels are used to train $\mathrm{Cls}_{\phi}(.)$ in step 12 for $\mathrm{epochs}_{\mathrm{cls}}$ steps and the corresponding checkpoints are saved. After the completion of iterations over $\tilde{\mathbf{b}}$, i.e., the steps 5--13, the metrics, e.g., accuracy and loss, are evaluated progressively using $eval_{\mathrm{metric},b}(.)$ for $b\in\tilde{\mathbf{b}}$ for the hold-out utterances in each $epoch$ in step 14. Subsequently, the parameters of early-stop$_{b}$, e.g., the internal counter, score, and status, are updated according to the progressive metrics from the previous step in step 15. In case the hardest progressive task, has a ``true'' early stopping status, the training will be stopped as described in steps 16--18. The hardest progressive task is the task after registering the last bucket $\tilde{\mathbf{b}}[-1]$ with the largest number of classes, the total number of speakers, and the fewest utterances per speaker, due to the limited allowed memory of $max_{\mathrm{mem}}$. Finally, the steps 2--19 are repeated for the specified range of $epochs$.

\subsection{Dynamic Registration of Speakers' Consent}

The dynamic registration of new speakers' consent to the pool of previously registered buckets of speakers is described in this subsection. For the dynamic process of registering new speakers' consent, it is required to optimally allocate the Euclidean space for new speakers. To this end, the shortest $L2$ pairwise distance is used as a metric to find the optimal buckets for new speakers. In other words, registering new speakers into the buckets with shortest $L2$ pairwise distance requires less Euclidean space. Consequently, it is possible to register more new speakers in the disjoint updated feature space of the buckets. This property is essential for bucket prediction that requires disjoint buckets in the feature space. According to the above explanations, the buckets with the shortest $L2$ pairwise distance from the new speakers are referred to as optimal buckets in this paper.

As the number of buckets is usually smaller than the number of new speaker registrations, there are at least two new speakers registered in the same bucket during each iteration. However, the registration of a new speaker in a bucket changes the contrastive feature state of that bucket such that it may no longer be the optimal bucket for registering the subsequent new speaker. As a result, new speakers in the subsequent round may select different optimal buckets, according to the shortest $L2$ pairwise distance, after registrations of new speakers in the current round. Consequently, the following function is applied to obtain the optimal buckets and the corresponding new speakers in the evaluation mode for each round, see the Appendix. \ref{app:opt-unique-bkts}. 
\begin{IEEEeqnarray}{rCl}
\widetilde{\mathbf{b}}^{*}_{\mathrm{reg}}, \widetilde{\mathbf{s}}^{*}_{\mathrm{reg}}, \mathbf{b}^{*}_{\mathrm{sofar}}, \mathbf{s}^{*}_{\mathrm{sofar}} &=& opt_{spk,bkt}(\bar{\mathbf{z}}^{\mathrm{eval}}, \tilde{\mathbf{b}}, \mathbf{s}_{\mathrm{reg}},
\widetilde{\mathbf{s}}^{u}_{\mathrm{reg}}, n_{\mathrm{round}}), \label{opt_spk_bkt}
\end{IEEEeqnarray}
where $\widetilde{\mathbf{b}}^{*}_{\mathrm{reg}}$ and $\widetilde{\mathbf{s}}^{*}_{\mathrm{reg}}$ denote unique optimal buckets and corresponding new speakers to be dynamically registered for the round $n_{\mathrm{round}}$. The terms $\mathbf{b}^{*}_{\mathrm{sofar}}$ and $\mathbf{s}^{*}_{\mathrm{sofar}}$ represent optimal buckets and corresponding new speakers that are already registered so far, i.e., prior to the round $n_{\mathrm{round}}$. In \eqref{opt_spk_bkt}, the function parameters $\bar{\mathbf{z}}^{\mathrm{eval}}$, $\mathbf{s}_{\mathrm{reg}}$, and $\widetilde{\mathbf{s}}^{u}_{\mathrm{reg}}$, denote a tuple of speaker embeddings in the evaluation mode according to the previously registered speakers and new speakers, the list of new speakers to be registered, and the list containing unique new speakers, respectively.

\begin{algorithm}[!t]\label{dyn_cnst_emb_rep}

\SetAlgoLined
\DontPrintSemicolon

\SetKwFunction{uoptbkt}{$opt_{spk,bkt}$}
\SetKwFunction{perroundsspksbkts}{$prop^{reg}_{spk,bkt}$}

\SetKwFunction{spattern}{$strategy\_pattern\_reg$}

\SetKwFunction{nutts}{$num_{spk,utts}$}
\SetKwFunction{ncollection}{$collection_{indx}$}
\SetKwFunction{interbkt}{$sample_{int-bkt}$}

\SetKwFunction{emetrics}{$eval_{\mathrm{metric},b}$}

Follow the steps in \eqref{opt_spk_bkt}-\eqref{prop_reg}, respectively.

Compute $\tilde{n}_{spk,\mathrm{utt}}$ according to \eqref{num_spk_utts}.

\For {$epoch$ in $range(epochs)$}{
\If{$len(\widetilde{\mathbf{b}}^{*}_{\mathrm{reg}})\: != \: 0$}{
Obtain $\mathcal{C}_{indx}$ according to \eqref{indx_collect}.\;
$\mathbf{zy}_{\mathrm{init}}$ $=$ ([], [])\;
\For {$\_$, $b$ in $enumerate(\tilde{\mathbf{b}})$}{

Load a random shard of old dataset for speakers in $b$ with $n_{utt}$ and $pcnt_{\mathrm{old}}$.

Load a random shard of new dataset according to $\mathcal{S}_{\mathrm{reg}}[b]$ with $n_{utt}$.

Combine the loaded old and new datasets from the previous steps.

\If{!(early-stop$_{b})$ \& $\mathcal{P}_{\mathrm{reg}}[b]=\mathrm{pattern}_{1/3}$}{
Train $\mathrm{Enc}_{\theta_{\mathrm{b}}}(.)$ for $\mathrm{epochs}_{\mathrm{cont}}$  contrastively using $\mathcal{P}_{\mathrm{reg}}[b]$ and save checkpoints.
}

Return the embeddings in \eqref{enc} for latest checkpoints and corresponding labels.
  
$\mathbf{D}^{max_{\mathrm{mem}}}_{\mathrm{buff}}$, $\mathbf{y}^{max_{\mathrm{mem}}}_{\mathrm{buff}}$ $=$ \interbkt{$\mathcal{C}_{indx}[b]$, $\mathbf{zy}_{b}$, $\mathbf{zy}_{\mathrm{init}}$}

Train $\mathrm{Cls}_{\phi}(.)$ using $\{\mathbf{D}^{max_{\mathrm{mem}}}_{\mathrm{buff}}, \mathbf{y}^{max_{\mathrm{mem}}}_{\mathrm{buff}}\}$ for $\mathrm{epochs}_{\mathrm{cls}}$ and save checkpoints.
}

Progressively evaluate \emetrics{.} for $b\in\tilde{\mathbf{b}}$.

Update ``early-stop$_{b}$'' parameters according to \emetrics{.}.

\If{early-stop$_{\tilde{\mathbf{b}}[-1]}$}{Break the training.}
}
}
\caption{Dynamic consent management for new speaker registrations}
\end{algorithm}

The process for the dynamic registration of new speaker(s) to the previously registered buckets of speakers is provided in the Algorithm. \ref{dyn_cnst_emb_rep}. For the initial round, i.e., the new registration round $n_{\mathrm{round}}=0$, $\widetilde{\mathbf{s}}^{u}_{\mathrm{reg}}$ is set to an empty list [], the list of new speakers to be registered $\mathbf{s}_{\mathrm{reg}}$ is initialized as $[N,\ldots, N + N_\mathrm{reg}-1]$ where $N$ and $N_{\mathrm{reg}}$ are number of old and new speakers, respectively. The list containing the number of speakers per buckets for dynamic registrations $\tilde{\mathbf{n}}_{bkt}$ is set to the initial state $\mathbf{n}_{bkt}$, i.e., a list containing the old number of speakers per buckets prior to dynamic registrations. The latest available checkpoints of $\mathrm{Enc}_{\{\theta_{\mathrm{b}}\}}(.)$ for $\forall b\in\tilde{\mathbf{b}}$, and $\mathrm{Cls}_{\phi}(.)$ in the current round are loaded. For $n_{\mathrm{round}} = 0$, the aforementioned checkpoints, except the last linear layer of the classifier with the output dimension of $N+N_{\mathrm{reg}}$, are loaded from the trained network with the Algorithm. \ref{cnst_emb_rep}    for the old $N$ speakers. Subsequently, the parameters for optimal buckets and corresponding new speakers are obtained as described in \eqref{opt_spk_bkt}. For the next round, $\mathbf{s}_{\mathrm{reg}}$ and $\widetilde{\mathbf{s}}^{u}_{\mathrm{reg}}$ are updated as follows.
\begin{IEEEeqnarray}{rCl}
&&
\mathbf{s}_{\mathrm{reg}}\gets\mathbf{s}_{\mathrm{reg}}\backslash\widetilde{\mathbf{s}}^{*}_{\mathrm{reg}} \quad \& \quad \widetilde{\mathbf{s}}^{u}_{\mathrm{reg}}\gets\widetilde{\mathbf{s}}^{*}_{\mathrm{reg}}. \label{next_round_set}
\end{IEEEeqnarray}
In other words, new speakers to be registered in the current round are excluded from $\mathbf{s}_{\mathrm{reg}}$ and $\widetilde{\mathbf{s}}^{u}_{\mathrm{reg}}$ is updated accordingly. The following function provides the necessary properties for registering new speakers, see the Appendix. \ref{app:prop-reg}.
\begin{IEEEeqnarray}{rCl}
\tilde{\mathbf{n}}_{bkt}, \tilde{\mathbf{n}}^{\mathrm{reg}}_{bkt}, \mathcal{S}_{\mathrm{reg}}, \mathcal{P}_{\mathrm{reg}} &=& prop^{reg}_{spk,bkt}(\tilde{\mathbf{b}}, 
\mathbf{n}_{bkt}, 
\widetilde{\mathbf{b}}^{*}_{\mathrm{reg}}, \mathbf{b}^{*}_{\mathrm{sofar}},  \widetilde{\mathbf{s}}^{*}_{\mathrm{reg}}, \mathbf{s}^{*}_{\mathrm{sofar}}), \label{prop_reg}
\end{IEEEeqnarray}
where $\tilde{\mathbf{n}}_{bkt}$ denotes the updated number of speakers per buckets, $\tilde{\mathbf{n}}^{\mathrm{reg}}_{bkt}$ is the updated number of new speakers per buckets containing the values of zero or one since at most one new speaker should be registered in each optimal bucket per round, $\mathcal{S}_{\mathrm{reg}}$ represents a dictionary of new speakers in buckets with the keys of $b\in\tilde{\mathbf{b}}$ and values of new speakers per buckets, and $\mathcal{P}_{\mathrm{reg}}$ denotes a dictionary of registration patterns in which the new speakers are registered in the buckets with the keys of $b\in\tilde{\mathbf{b}}$ and values of pattern status. After step 1 and obtaining the required parameters as described in \eqref{opt_spk_bkt}-\eqref{prop_reg}, respectively, the updated number of utterances per speakers $\tilde{n}_{spk,\mathrm{utt}}$ are obtained according to the updated number of speakers per buckets $\tilde{\mathbf{n}}_{bkt}$, the updated number of new registrations per buckets $\tilde{\mathbf{n}}^{\mathrm{reg}}_{bkt}$, and the maximum allowed memory $max_{\mathrm{mem}}$, based on \eqref{num_spk_utts} in step 2.

The training iterations over the specified range of $epochs$ starts at step 3. Prior to starting the registration of the new speakers in the buckets $\tilde{\mathbf{b}}$, the algorithm checks if the unique set of optimal buckets is non-empty in step 4. Using the computed $\tilde{n}_{spk,\mathrm{utt}}$ in step 2, $\tilde{\mathbf{n}}_{bkt}$, and $\tilde{\mathbf{n}}^{\mathrm{reg}}_{bkt}$, a flattened collection of indices per speakers in buckets are obtained, according to \eqref{indx_collect} in step 5. A tuple of speaker embeddings and corresponding labels are set to empty lists in step 6. The iterations over $\tilde{\mathbf{b}}$ starts in step 7 where the $enumerate(.)$ generates a specific bucket $b$ for each iteration. In step 8, a random shard of $pcnt_{\mathrm{old}}\%$ of old dataset for $n_{utt}$ utterances per speakers in $b$ is loaded. The term $pcnt_{\mathrm{old}}\%$ denotes the total percentage of utterances of the old speakers, previously registered using Algorithm. \ref{cnst_emb_rep}. Subsequently, a random shard of new dataset for $n_{utt}$ utterances according to the new speakers in $b$, $\mathcal{S}_{\mathrm{reg}}[b]$, is loaded in step 9. The loaded datasets from the previous steps are combined in step 10. In case the early stopping obtained according to the progressive evaluation up to bucket $b$ for a given metric, e.g., accuracy or loss, is not true, shown as !(early-stop$_{b}$), and the registration pattern follows $\mathrm{pattern}_{1/3}$, train $\mathrm{Enc}_{\theta_{\mathrm{b}}}(.)$ for $\mathrm{epochs}_{\mathrm{cont}}$ steps contrastively according to the given pattern shown in steps 11--13. As the $\mathrm{pattern}_{1/3}$ represents registration of the new speaker(s) in the corresponding optimal bucket(s), it requires training of the contrastive feature encoder accordingly. However, the $\mathrm{pattern}_{2/4}$ does not require training of the contrastive feature encoder, as it represents the already registered new speakers, $\mathrm{pattern}_{2}$, or previously registered old speakers, $\mathrm{pattern}_{4}$, Appendix. \ref{app:prop-reg}.

Next, the speaker embeddings are obtained according to \eqref{enc} for the latest available checkpoints and corresponding labels are returned for $b$ in step 14. Using the inter-bucket sampling function $sample_{int-bkt}(.)$, the progressive features $\mathbf{D}^{max_{\mathrm{mem}}}_{\mathrm{buff}}$ and corresponding labels $\mathbf{y}^{max_{\mathrm{mem}}}_{\mathrm{buff}}$ are obtained with the memory size of $max_{\mathrm{mem}}$ in step 15. The progressive features and the corresponding labels are used to train $\mathrm{Cls}_{\phi}(.)$ in step 16 for $\mathrm{epochs}_{\mathrm{cls}}$ steps and the checkpoints are saved. After the completion of iterations over $\tilde{\mathbf{b}}$, i.e., the steps 7--17, the metrics, e.g., accuracy and loss, are evaluated progressively using $eval_{\mathrm{metric},b}(.)$ for $b\in\tilde{\mathbf{b}}$ for the hold-out utterances in each $epoch$ in step 18. Consequently, the parameters of early-stop$_{b}$ are updated according to the progressive metrics from the previous step in step 19. In case the hardest progressive task, with the same definition as in the Algorithm. \ref{cnst_emb_rep}, has a true early stopping status, the training will be stopped as described in steps 20--22. Finally, the steps 3--24 are repeated for the specified range of $epochs$ in each round.

\subsection{Dynamic Removal of Speakers' Consent}

The process for removing the previously registered speakers from the buckets is proposed in the Algorithm. \ref{unreg_rereg_cnst_emb_rep}. The parameters of $\mathrm{Enc}_{\{\theta_{\mathrm{b}}\}}(.)$, $\{\theta_{\mathrm{b}}\}$, are initialized based on the checkpoints of previously registered speakers for $b\notin\tilde{\mathbf{b}}_{\mathrm{unreg}}$ and the available checkpoint of the remaining speakers for $b\in\tilde{\mathbf{b}}_{\mathrm{unreg}}$. Accordingly, the parameters of $\mathrm{Cls}_{\phi}(.)$, $\phi$, are initialized based on latest available checkpoints. As there are no new speakers to be registered, $\mathbf{n}^{\mathrm{reg}}_{bkt}$ is set to zero. The properties of interest for removing the given set of speakers from the pool of previously registered speakers are obtained in step 1 as follows.
\begin{IEEEeqnarray}{rCl}
\tilde{\mathbf{n}}_{bkt}, \mathcal{S}_{\mathrm{res}}, \mathcal{P}_{\mathrm{unreg}} &=& prop^{unreg}_{spk,bkt}(\tilde{\mathbf{b}}, \mathbf{n}_{bkt}, \tilde{\mathbf{b}}_{\mathrm{unreg}}, \mathbf{s}_{\mathrm{res}}), \label{prop_unreg}
\end{IEEEeqnarray}
where $\tilde{\mathbf{b}}_{\mathrm{unreg}}$ denotes the corresponding unique set of buckets for removing the given set of speakers, and $\mathbf{s}_{\mathrm{res}}$ represents the set of residual speakers after removing the given set of speakers. Consequently, the properties of interest, including the updated number of speakers per buckets $\tilde{\mathbf{n}}_{bkt}$; a dictionary of updated residual speakers in buckets $\mathcal{S}_{\mathrm{res}}$ with the keys of $b\in\tilde{\mathbf{b}}$ and values of residual speakers per buckets; and a dictionary of patterns for removing the speakers with the keys of $b\in\tilde{\mathbf{b}}$ and values of pattern status, are obtained using the function $prop^{unreg}_{spk,bkt}(.)$, see the Appendix. \ref{app:prop-unreg}. The proposed algorithm assumes the existence of at least two residual speakers per bucket required for contrastive training. In the results, it is explained how to deal with other cases. Applying the updated number of speakers per buckets $\tilde{\mathbf{n}}_{bkt}$ and maximum allowed memory $max_{\mathrm{mem}}$, the number of utterances per speaker $\tilde{n}_{spk,\mathrm{utt}}$ is computed according to \eqref{num_spk_utts} in step 2. 

\RestyleAlgo{ruled}
\begin{algorithm}[!t]\label{unreg_rereg_cnst_emb_rep}

\SetAlgoLined
\DontPrintSemicolon

\SetKwFunction{unregspksbkts}{$prop^{unreg}_{spk,bkt}$}

\SetKwFunction{spattern}{$strategy\_pattern\_reg$}

\SetKwFunction{nutts}{$num_{spk,utts}$}
\SetKwFunction{ncollection}{$collection_{indx}$}
\SetKwFunction{interbkt}{$sample_{int-bkt}$}

\SetKwFunction{emetrics}{$eval_{\mathrm{metric},b}$}

Compute the parameters based on \eqref{prop_unreg}.

Compute $\tilde{n}_{spk,\mathrm{utt}}$ according to \eqref{num_spk_utts}.

\For {$epoch$ in $range(epochs)$}{

Obtain $\mathcal{C}_{indx}$ according to \eqref{indx_collect}.\;

$\mathbf{zy}_{\mathrm{init}}$ $=$ ([], [])\;
\For {$\_$, $b$ in $enumerate(\tilde{\mathbf{b}})$}{

Load a random shard of dataset according to $\mathcal{S}_{\mathrm{res}}[b]$ with $n_{utt}$.

\If{!(early-stop$_{b})$ \& $\mathcal{P}_{\mathrm{unreg}}[b]=\mathrm{pattern}_{1}$}{
Train $\mathrm{Enc}_{\theta_{\mathrm{b}}}(.)$ for $\mathrm{epochs}_{\mathrm{cont}}$  contrastively using $\mathcal{P}_{\mathrm{unreg}}[b]$ and save checkpoints.
}

Return the embeddings in \eqref{enc} for latest checkpoints and corresponding labels.
  
$\mathbf{D}^{max_{\mathrm{mem}}}_{\mathrm{buff}}$, $\mathbf{y}^{max_{\mathrm{mem}}}_{\mathrm{buff}}$ $=$ \interbkt{$\mathcal{C}_{indx}[b]$, $\mathbf{zy}_{b}$, $\mathbf{zy}_{\mathrm{init}}$}

Train $\mathrm{Cls}_{\phi}(.)$ using $\{\mathbf{D}^{max_{\mathrm{mem}}}_{\mathrm{buff}}, \mathbf{y}^{max_{\mathrm{mem}}}_{\mathrm{buff}}\}$ for $\mathrm{epochs}_{\mathrm{cls}}$ and save checkpoints.
}

Progressively evaluate \emetrics{.} for $b\in\tilde{\mathbf{b}}\backslash\tilde{\mathbf{b}}_{\mathrm{unreg}}$, and per bucket for $b\in\tilde{\mathbf{b}}_{\mathrm{unreg}}$.
  
Update ``early-stop$_{b}$'' parameters according to \emetrics{.}.

\If{early-stop$_{\tilde{\mathbf{b}}\backslash\tilde{\mathbf{b}}_{\mathrm{unreg}}[-1]}$ and $all$(early-stop$_{b\in\tilde{\mathbf{b}}_{\mathrm{unreg}}}$)}{Break the training.}

}
\caption{Consent management for removing previously registered speakers}
\end{algorithm}

The training iterations over the specified range of $epochs$ starts at step 3. Prior to starting the removal of the speakers from the buckets $\tilde{\mathbf{b}}$, $\mathcal{C}_{indx}$ is obtained according to \eqref{indx_collect} using $\tilde{n}_{spk,\mathrm{utt}}$ in step 4. A tuple of speaker embeddings and corresponding labels are set to empty lists in step 5. The iterations over $\tilde{\mathbf{b}}$ starts in step 6 where the $enumerate(.)$ generates a specific bucket $b$ for each iteration. In step 7, a random shard of dataset for $n_{utt}$ utterances per speakers according to $\mathcal{S}_{\mathrm{res}}[b]$ for each $epoch$ is loaded. The early stopping status is obtained according to the progressive evaluation of all the buckets except the unique set of buckets for removing $\tilde{\mathbf{b}}_{\mathrm{unreg}}$, shown as $b\in\tilde{\mathbf{b}}\backslash\tilde{\mathbf{b}}_{\mathrm{unreg}}$, and per bucket evaluation of $b\in\tilde{\mathbf{b}}_{\mathrm{unreg}}$ for a given metric, e.g., accuracy or loss. This is due to the fact that evaluation metric for the bucket(s) comprising the unregistered speaker(s) is obtained for the entire hold-out utterances including the unregistered speakers. For example, if one speaker is removed from a given bucket with $5$ speakers, the expected metric for accuracy of that bucket is around $80\%$. Consequently, if the early stopping status is not true, i.e., !(early-stop$_{b}$), and the removal pattern follows $\mathrm{pattern}_{1}$, then $\mathrm{Enc}_{\theta_{\mathrm{b}}}(.)$ is trained for $\mathrm{epochs}_{\mathrm{cont}}$ steps contrastively according to the given pattern, steps 8--10. This is due to the fact that $\mathrm{pattern}_{1}$ requires training of the contrastive features excluding the samples of the unregistered speaker(s) from the bucket for selective forgetting. On the other hand, $\mathrm{pattern}_{2}$ does not require training of the contrastive features as it is related to the bucket(s) that do not include unregistered speaker(s), see the Appendix. \ref{app:prop-unreg}.

Next, the speaker embeddings are obtained according to \eqref{enc} for the latest available checkpoints and corresponding labels are returned in step 11. The progressive features $\mathbf{D}^{max_{\mathrm{mem}}}_{\mathrm{buff}}$ and corresponding labels $\mathbf{y}^{max_{\mathrm{mem}}}_{\mathrm{buff}}$ with $max_{\mathrm{mem}}$ are obtained in step 12, and used to train $\mathrm{Cls}_{\phi}(.)$ in step 13 for $\mathrm{epochs}_{\mathrm{cls}}$ steps. After the completion of iterations over $\tilde{\mathbf{b}}$, i.e., the steps 6--14, the metrics, e.g., accuracy and loss, are evaluated progressively using $eval_{\mathrm{metric},b}(.)$ for $b\in\tilde{\mathbf{b}}\backslash\tilde{\mathbf{b}}_{\mathrm{unreg}}$ and per bucket for $b\in\tilde{\mathbf{b}}_{\mathrm{unreg}}$ for the hold-out utterances in each $epoch$ in step 15. Consequently, the parameters of early-stop$_{b}$ are updated according to the metrics from the previous step in steps 16. In case the hardest progressive task, excluding the $\tilde{\mathbf{b}}_{\mathrm{unreg}}$ and shown as early-stop$_{\tilde{\mathbf{b}}\backslash\tilde{\mathbf{b}}_{\mathrm{unreg}}[-1]}$, with the same definition as in the Algorithm. \ref{cnst_emb_rep}, together with all the bucket(s) comprising the unregistered speaker(s), $all$(early-stop$_{b\in\tilde{\mathbf{b}}_{\mathrm{unreg}}}$), have true early stopping statuses, the training will be stopped as described in steps 17--19. Finally, the steps 3--20 are repeated for the specified range of $epochs$. The process for re-registering follows a similar procedure as in Algorithm. \ref{unreg_rereg_cnst_emb_rep} by re-registering the unregistered speaker(s) in the corresponding bucket(s).

It is worth noting that the bucket index may encode information about the duration of dissent in practice. This way, speakers that do not provide consent for a given time interval are grouped in the buckets with the corresponding time stamps stored as a decision tree in the back-end. Consequently, the problem boils down to a decomposable search algorithm that is known to be a fully retro-active data structure via decision trees with the overhead of $O(\log(B))$ for $B$ buckets \cite{demaine2007retroactive}. 


\section{Experiments} \label{results}

The goal of the simulations is to answer the following questions for both supervised and unsupervised modes:
\begin{enumerate}
\item Can the proposed method enable a fast training?
\item Can the proposed method dynamically register new speakers efficiently?
\item Can the proposed method dynamically remove and re-register the speakers efficiently?
\item Can the proposed method provide a good verification performance?
\end{enumerate}
All the experiments were run on a single NVIDIA GeForce RTX 2070 GPU and Python version 3.9.4 was used to implement the algorithms. The code for the simulations will be made available.
\subsection{Dataset}

The LibriSpeech\footnote{The main reason for using this dataset is to provide free access to the results and reproducibility of the simulations in terms of both source code and dataset. The interested researchers are welcome to extend the results for other datasets, e.g., NIST SRE evaluation campaigns \cite{912681}, TIMIT \cite{AB2/SWVENO_1993}, and so on.} dataset is used for all the results \cite{Librispeech}. Different subsets of the aforementioned dataset is used for training and testing. In particular, $5$ agents are used for the simulations each of which using $N=40$ different speakers selected from the set of speakers with lower word error rate, denoted by ``clean'' in the LibriSpeech dataset, according to the agent index. In other words, for Agent $i$, speakers $[i\times N, (i+1)\times N)$ are selected and equally divided in $B=8$ different buckets. For the registration of new speakers in the previously trained contrastive buckets of speakers, $N_{\mathrm{reg}}=20$ speakers are selected from the set of speakers with higher word error rate, denoted by ``other'' in the LibriSpeech dataset and briefly referred to as new speakers with noisy utterances, according to the agent index. In other words, for Agent $i$, new speakers $[i\times N_{\mathrm{reg}}+N, (i+1)\times N_{\mathrm{reg}}+N)$ are dynamically registered in the pool of previously registered speakers $[i\times N, (i+1)\times N)$.

\subsection{Hyper-Parameters and Network Architecture}

\begin{table}[!t]
\renewcommand{\arraystretch}{1}
\caption{Feature Extraction Network per Bucket}
\label{feat_extract_netw_arch}
\centering
  \begin{tabular}{lSSSSSS}
    \midrule
	
    \multirow{1}{*}{\#} &
    \multicolumn{1}{c}{Layer (Type)} &
      \multicolumn{1}{c}{Output Shape} &
      \multicolumn{1}{c}{Param \#} 
      \\
      \midrule
    1 & \text{LSTM} & \text{[-1, 160, 128]} & \text{351,232} \\
    2 & \text{Linear} & \text{[-1, 160, 256]} & \text{33,024} \\
    3 & \text{tanh} & \text{[-1, 160, 256]} & \text{0} \\
    4 & \text{GroupNorm} & \text{[-1, 160, 256]} & \text{320} \\
    5 & \text{Attention Pooling} & \text{[-1, 256]} & \text{257} \\
    6 & \text{Normalization} & \text{[-1, 256]} & \text{0} \\
    \bottomrule
  \end{tabular}
\end{table}

\begin{table}[!t]
\renewcommand{\arraystretch}{1}
\caption{Classifier Network}
\label{classifier_netw_arch}
\centering
  \begin{tabular}{lSSSSSS}
    \midrule
	
    \multirow{1}{*}{\#} &
    \multicolumn{1}{c}{Layer (Type)} &
      \multicolumn{1}{c}{Output Shape} &
      \multicolumn{1}{c}{Param \#} 
      \\
      \midrule
    1 & \text{Linear} & \text{[-1, 64]} & \text{16,448} \\
    2 & \text{ReLU} & \text{[-1, 64]} & \text{0} \\
    3 & \text{Linear} & \text{[-1, 64]} & \text{4,160} \\
    4 & \text{ReLU} & \text{[-1, 64]} & \text{0} \\
    5 & \text{Linear} & \text{[-1, N]} & \text{2,600} \\
    6 & \text{Softmax} & \text{[-1, N]} & \text{0} \\
    \bottomrule
  \end{tabular}
\end{table}

The log \ac{MFB} features with the feature dimension of $40$, the frame length of $25$ \ac{ms}, the stride of $10$ \ac{ms}, and the \ac{VAD} of $20$ dB are used as the input features $\mathbf{x}_{s,b}$ for the encoder $\mathrm{Enc}_{\theta_{b}}(.)$ in \eqref{enc}. Subsequently, the features are normalized and scaled by the mean and variance, respectively, along the time-axis. Finally, the number of iterations for the contrastive feature extraction is set to $\mathrm{epochs}_{\mathrm{cont}}=5$ and the number of iterations for the classifier are set to $\mathrm{epochs}_{\mathrm{cls}}=2$ and $\mathrm{epochs}_{\mathrm{cls}}=1$ for the supervised and unsupervised cases, respectively, as they provide optimal performance in terms of total elapsed time for training.

The per bucket embedding network $\mathrm{Emb}_{\theta_{e,b}}(.)$ in \eqref{enc} is implemented according to Table. \ref{feat_extract_netw_arch}, where $-1$ in the output shape column denotes the batch dimension of a tensor. In particular, the \ac{LSTM} layer is applied with feature dimension $40$, cell dimension $128$, and number of layers $3$. The group-norm layer with the number of groups $4$ and the number of channels set to the segmentation length of $160$ is used according to \cite{Wu_2018_ECCV}. To obtain the attention weights in the projection head in \eqref{enc} required for the attentive pooling, the linear transformation with the input dimension $256$ and output dimension $1$ with the Softmax activation is applied. Subsequently, the embedding terms $\mathbf{z}_{s,b}$ in \eqref{enc} are obtained by multiplication of the attention weights from the previous step with $\mathrm{Emb}_{\theta_{e,b}}(\mathbf{x}_{s,b})$, summation over the segmentation length, and normalizing by the Euclidean norm over the embedding dimension. The classifier $\mathrm{Cls}_{\phi}(.)$ is implemented according to Table. \ref{classifier_netw_arch}. For dynamic registrations, $N$ is replaced by $N+N_{\mathrm{reg}}$ in the last Linear layer. The \ac{SGD} and \ac{Adam} optimizers are used for the supervised contrastive learning and classifications, respectively.

The same embedding network architecture was used to implement the algorithms for the unsupervised case. For the classification using the unsupervised learning, the first two layers of $\mathrm{Cls}_{\phi}(.)$ were used with the same hyper-parameters and the output layer was removed. For the unsupervised contrastive learning both the embedding network and the latent feature classification are optimized using \ac{SGD} using contrastive unsupervised learning for $\mathrm{epochs}_{\mathrm{cont}}=5$ and $\mathrm{epochs}_{\mathrm{cls}}=1$ iterations, respectively. It is worth noting that the performance, in terms of accuracy for the unsupervised learning, is obtained according to the ``cosine similarity matrix'' of the output features.  

\subsection{Baseline}

The performance of the proposed approach is compared to the baseline method applying unsupervised contrastive learning \cite{ICASSP_GE2E}. The main purpose of providing a baseline is not about comparing different architectures for contrastive learning, but to observe the effects of training with the proposed algorithms. In other words, the effects of the proposed methods on the training elapsed time, sample efficiency, and performance are of particular interest. Consequently, contrastive learning based methods with different architectures can benefit from the proposed algorithms in terms of convergence speed, efficient sampling, and dynamic capabilities \cite{Sang2022,Zhang2021ContrastiveSL,9414973}. For all the results, the baseline and the proposed method follow a similar network structure for speaker embedding. The hyper-parameters for the baseline are selected to provide comparable performance. In particular, the number of hidden nodes is set to $512$ with the projection size of $256$ and the $3$-layer \ac{LSTM} as in \cite{ICASSP_GE2E}.

\subsection{Results}

\begin{figure}[t]
    \centering
    \includegraphics[scale=0.8]{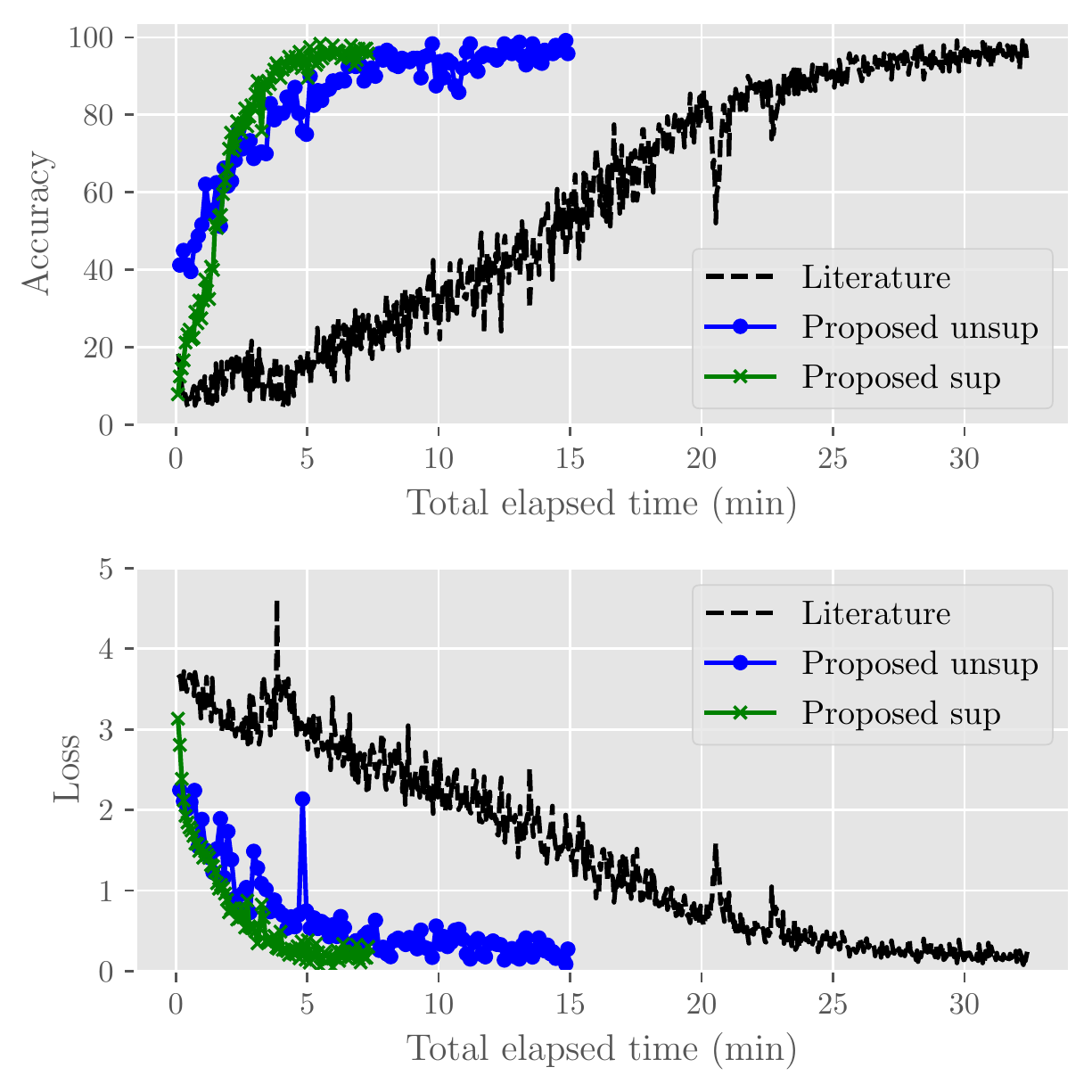}
    \caption{The comparison between testing accuracies and losses of an agent using the proposed contrastive embedding replay, with multi-strided progressive sampling in supervised and unsupervised modes, and the baseline method from the literature with respect to the elapsed time for training.}
    \label{acc_loss_sup_unsup}
\end{figure}

Fig. \ref{acc_loss_sup_unsup} shows the performance during testing in terms of accuracies and losses, for a given agent, using the proposed method in supervised and unsupervised modes with respect to the total elapsed time for training. It is observed that the proposed method in the supervised and unsupervised modes requires approximately $7$ and $15$ minutes to break the training loop by activating the early stopping mechanism in Algorithm. \ref{cnst_emb_rep}. However, the method from the literature requires approximately $32$ minutes to provide a similar performance. Moreover, the baseline method requires large batch size of $20\times N$ for $N=40$ speakers during each iteration compared to the proposed approach that only requires $max_{\mathrm{mem}}=120$. Consequently, the proposed method provides efficient use of data due to:
\begin{enumerate}
\item Dividing different sets of speakers in the buckets
\item Contrastive learning of speaker equivariance inductive bias only a few steps into each training iteration
\item Progressive increase of task difficulty by increasing number of speakers and decreasing number of utterances per speaker for a given memory budget
\item Per bucket early stopping of contrastive feature training based on the progressive evaluation of a given metric
\end{enumerate}

\begin{figure}[t]
    \centering
    \includegraphics[scale=0.8]{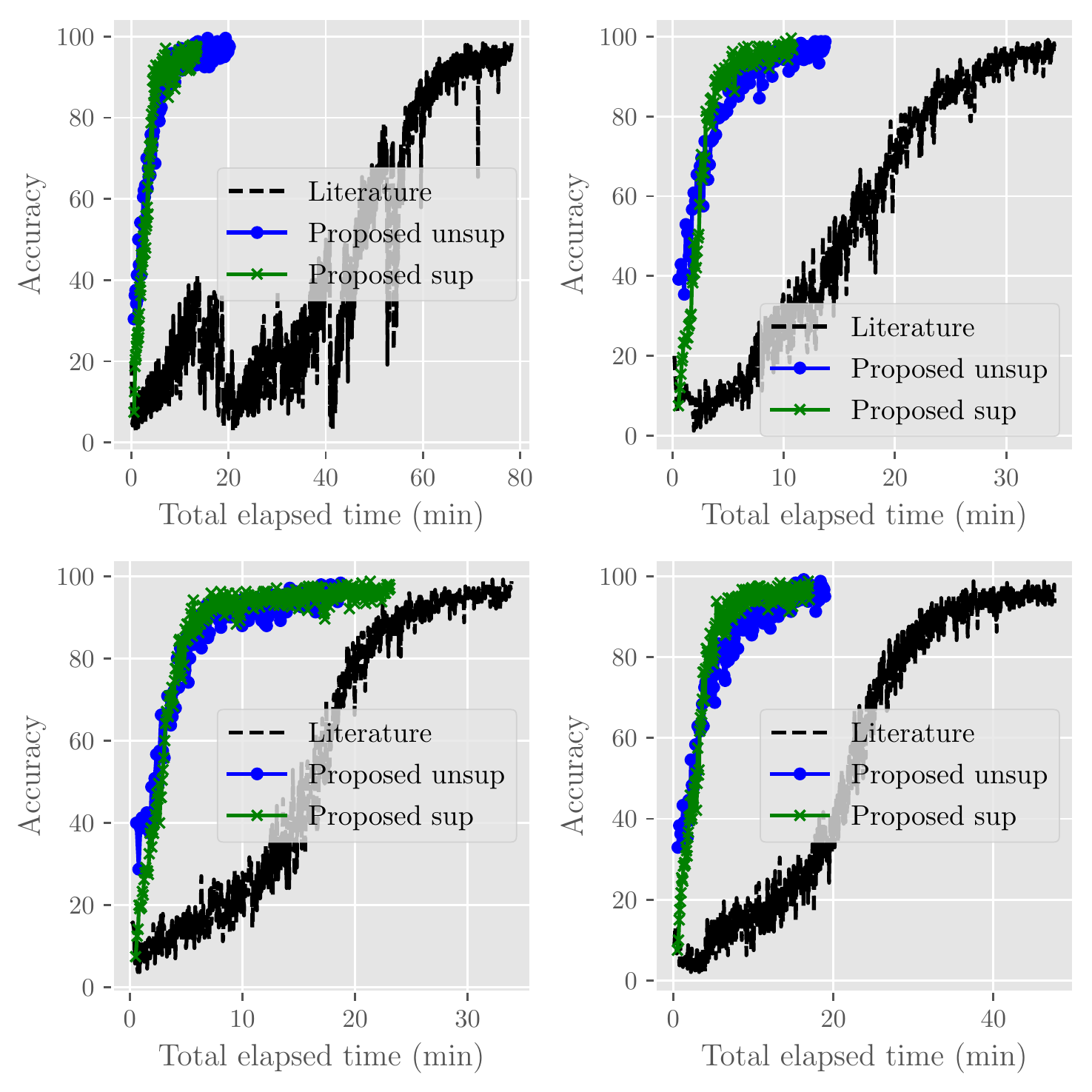}
    \caption{The comparison between testing accuracies of different agents using the proposed contrastive embedding replay, with multi-strided progressive sampling in supervised and unsupervised modes, and the baseline method from the literature with respect to the elapsed time for training.}
    \label{acc_sup_unsup}
\end{figure}

Fig. \ref{acc_sup_unsup} shows the performance during testing in terms of accuracies for different agents, using the proposed method in supervised and unsupervised modes, together with the method from the literature, with respect to the total elapsed time for training. It is observed that for all the agents the proposed method breaks the training loop, by satisfying the early stopping mechanism in Algorithm. \ref{cnst_emb_rep}, much faster than the method from the literature. In particular, the method from the literature approximately requires $\{78,34,34,48\}$ minutes to complete the training, with the similar performance, for the top to bottom plots and from left to right, respectively. However, the corresponding values using the proposed method are approximately $\{13,11,23,17\}$ minutes and $\{20,14,19,19\}$ minutes for supervised and unsupervised modes, respectively. Also, the baseline model with the aforementioned hyper-parameters requires orders of magnitude more parameters for speaker embedding per agent to provide a comparable performance. This is mainly related to the large batch size requirements and inefficient use of speaker equivariance inductive bias provided by contrastive features during the training. In conclusion, the proposed approach converges much faster than the baseline due to the aforementioned four points.

\begin{figure}[t]
    \centering
    \includegraphics[scale=0.8]{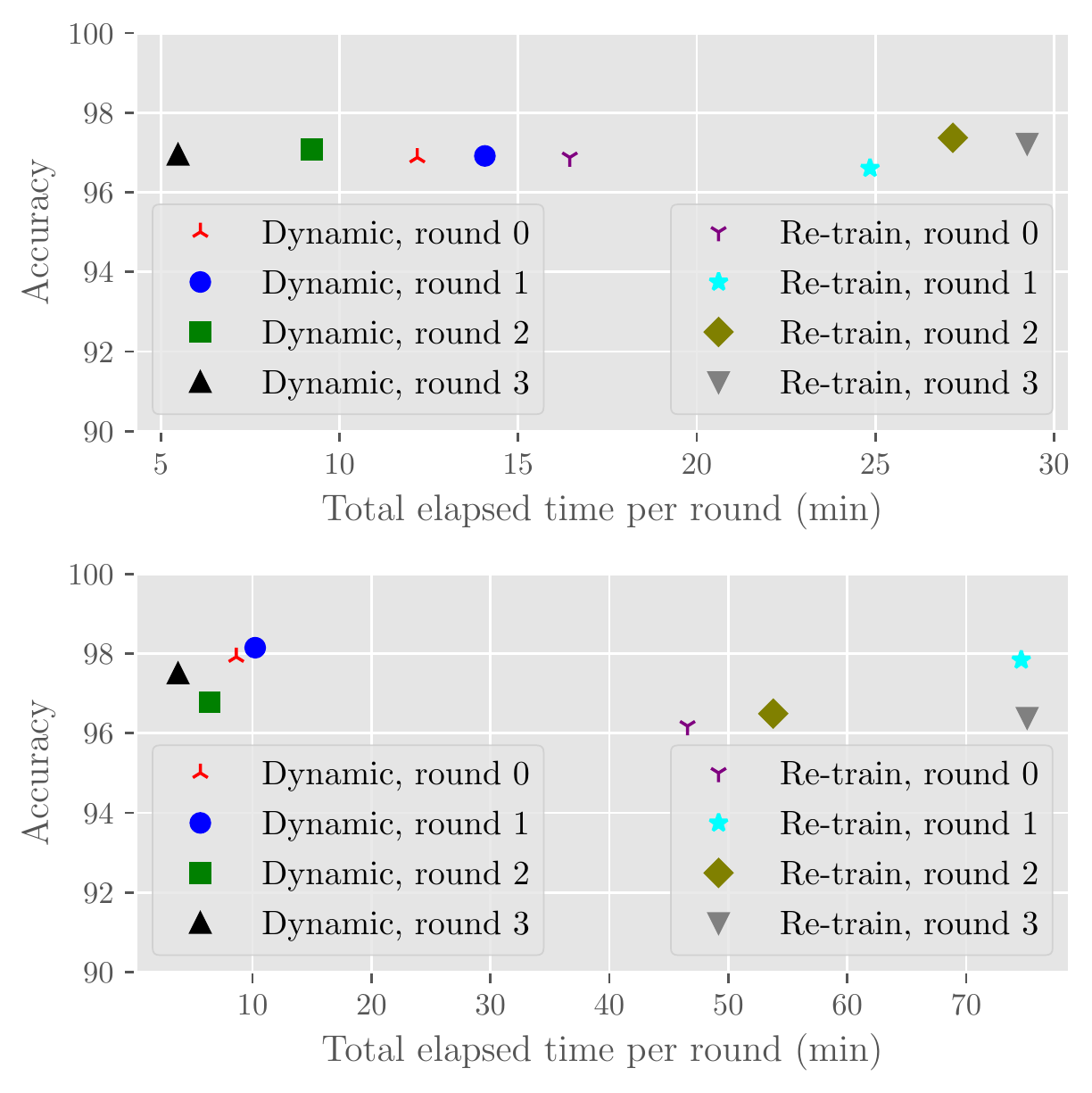}
    \caption{The testing accuracies per round for dynamic (top) supervised and (bottom) unsupervised registrations with respect to required elapsed time to break the registration loop. Different markers and colors are used to distinguish between different rounds of dynamic registrations. The corresponding values by re-training the network per rounds are reported by different markers.}
    \label{acc_reg_sup_elapsed_time}
\end{figure}

Fig. \ref{acc_reg_sup_elapsed_time} shows testing accuracy, for a given agent, with respect to the required elapsed time to break the dynamic registration training loop using the proposed method in the Algorithm. \ref{dyn_cnst_emb_rep} for different rounds of registrations. Only $50\%$ of the utterances of the previously registered $40$ speakers are used for this simulation. The performance for different percentage of old utterances for previously registered speakers are provided subsequently. The testing accuracy of each round is reported with respect to the registration elapsed time, i.e., after breaking the training by satisfying the early stopping condition in Algorithm. \ref{dyn_cnst_emb_rep} for each round. The result is reported using different markers and colors for different rounds of registrations. The corresponding values by re-training the network during each round, using the Algorithm. \ref{cnst_emb_rep} for full utterances of old and new speakers and without the dynamic registration mechanism in the Algorithm. \ref{dyn_cnst_emb_rep}, are reported by different colors and markers. It is observed that the proposed dynamic registration method provides much faster registrations compared to re-training the network for each round for both supervised and unsupervised cases. In particular, by increasing the number of rounds, the total number of speakers is increased; however, due to the efficient mechanism for dynamic registrations using the information from the previous rounds of registrations, the elapsed time for subsequent registrations is decreased. On the other hand, the elapsed time of re-training the network by increasing the number of rounds is increased due to increase in the overall number of speakers. For instance, the required elapsed times, to break the training loop, at the the end of dynamic registration rounds, i.e., round $3$, are approximately $\{5.5, 3.8\}$ minutes while it requires re-training the network for approximately $\{29, 75\}$ minutes for supervised and unsupervised cases, respectively.

\begin{figure}[!t]
    \centering
    \includegraphics[scale=0.8]{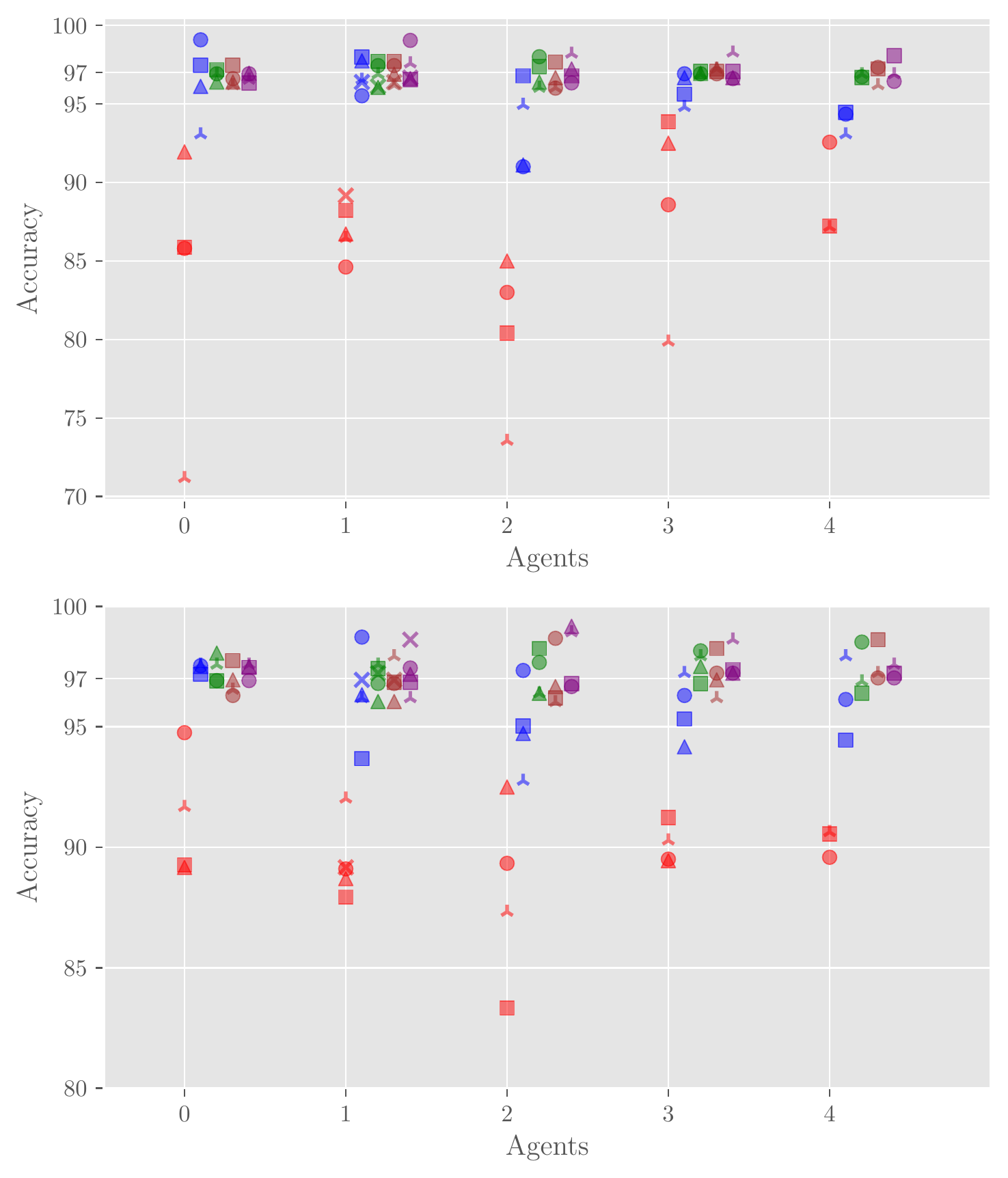}
    \caption{The testing accuracies per round with the different percentages of old utterances for different agents in the case of dynamic (top) supervised and (bottom) unsupervised registrations. Different markers denote different rounds, and the $\{10, 30, 50, 70, 90\}\%$ percentages of old utterances are color coded for each agent from left to right, respectively. For visibility purposes, the values using different percentages of old utterances are slightly shifted to the right for each agent.}
    \label{acc_reg_utts_pcnt}
\end{figure}

Fig.\ref{acc_reg_utts_pcnt} shows the testing accuracy per round for different percentage of old speaker utterances for agents 0--4, using the proposed dynamic consent management algorithm for registering new speakers in the supervised, top plot, and unsupervised, bottom plot, modes. The $\{10, 30, 50, 70, 90\}\%$ percentages of old utterances are color coded for each agent from left to right, respectively. Moreover, different rounds of registrations are shown using different markers. It is observed that for all the agents using $pcnt\geq 50\%$ of old utterances provides the required condition for breaking the dynamic registration loop in Algorithm. \ref{dyn_cnst_emb_rep} in the supervised and unsupervised modes. However, for some agents even using $pcnt= 30\%$ of old utterances is enough to provide a similar performance, e.g., agents $\{0, 1, 3\}$ in the supervised mode and agent $\{0\}$ in the unsupervised mode. The rest of agents using $pcnt= 30\%$ of old utterances provide a relatively good performance for both supervised and unsupervised modes; however their performance is slightly degraded in certain rounds. The performance starts to degrade by using only $pcnt = 10\%$ of old utterances for all the agents and all rounds. This is in particular due to the parameter shift towards the new speaker utterances. Using a portion of old utterances during dynamic registrations is extremely useful as the old utterances are not kept unnecessarily in the back-end during new registrations, hence providing improved privacy. In other words, the proposed dynamic registration strategy provides efficient use of data from the old speakers such that only a portion of the old utterances are required during the registrations of the new speakers without sacrificing the performance leading to improved privacy. Finally, it is possible to apply different hyper-parameter optimization, and choose different values for metrics to update early stopping counter and achieve higher testing accuracy. These points are not the main purpose of this work.

\begin{figure}[!t]
    \centering
    \includegraphics[scale=0.8]{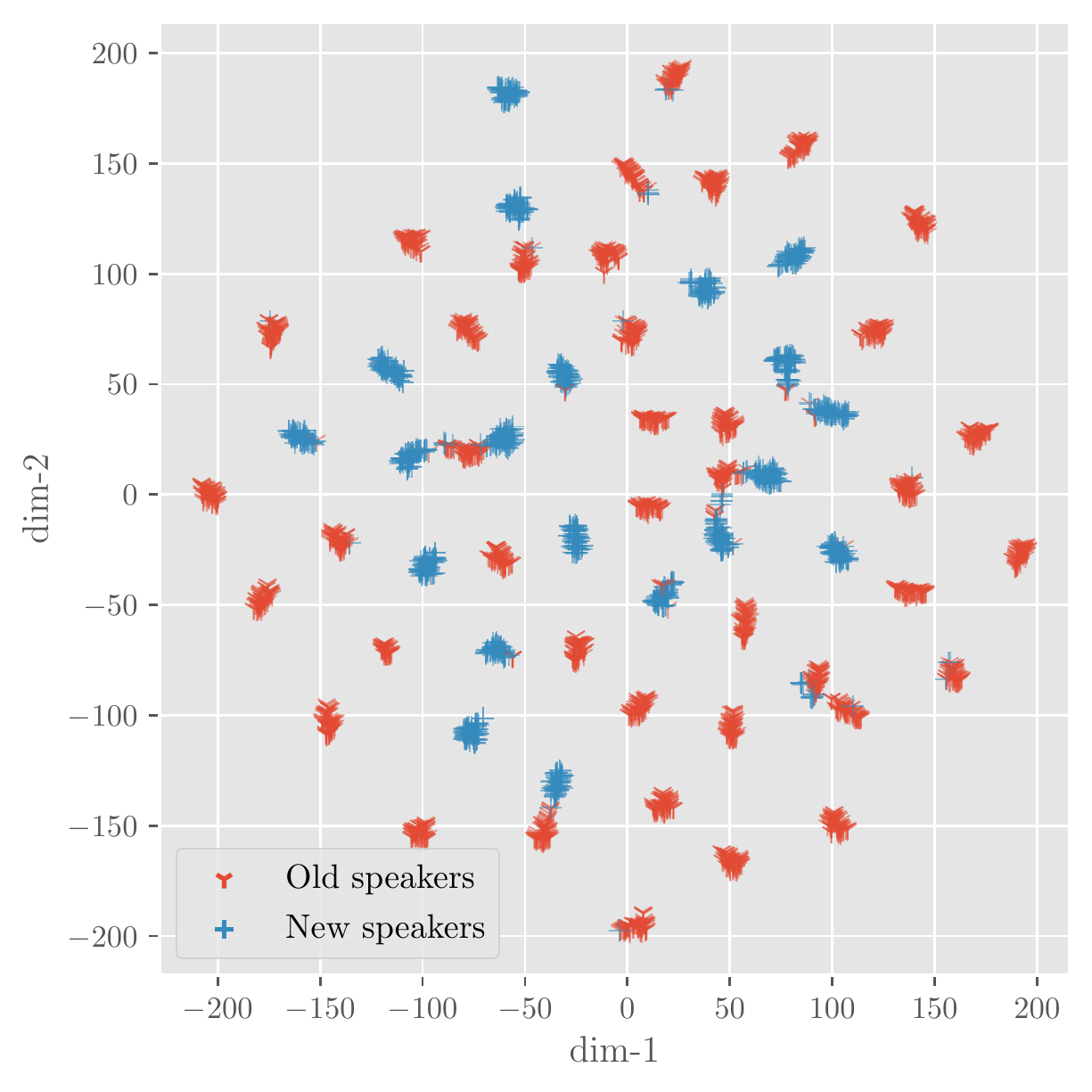}
    \caption{Visualization of the trained latent features after the second linear layer of the classifier during testing using t-SNE. The old previously registered $40$ speakers and the new dynamically registered $20$ speakers are shown with different markers and colors.}
    \label{dyn_reg_tsne}
\end{figure}

Fig. \ref{dyn_reg_tsne} shows the \ac{t-SNE} for the dynamically trained latent features after the second linear layer of the classifier during the testing \cite{vanDerMaaten2008}. It is observed that the separation between latent features of different speakers is almost perfect for the old speakers, new registrations, and among old and new features as shown by different colors and markers. In particular, the new registrations are distributed in different regions of the Euclidean space, and they are separable from the old speakers in different buckets. It is worth mentioning that the number of new registrations for each agent is upper bounded according to the limitations imposed by the Euclidean space, i.e., it is not possible to register arbitrarily large number of new speakers in each agent. Consequently, it is recommended to either create new agents or distribute new registrations between multiple agents in this case.

\begin{figure}[!t]
    \centering
    \includegraphics[scale=0.8]{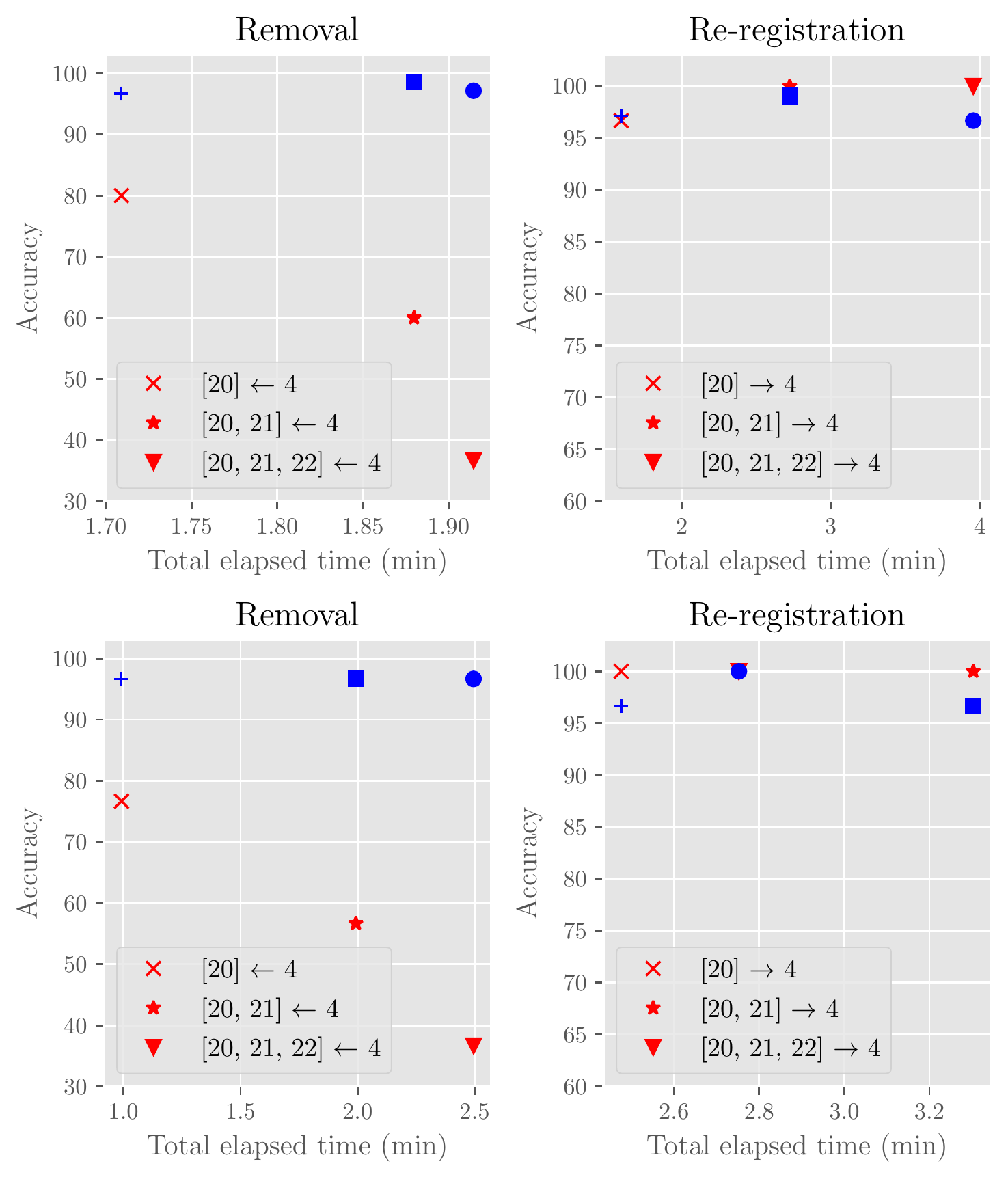}
    \caption{The testing accuracies with respect to total elapsed time per removal/re-registration for (top) supervised and (bottom) unsupervised. The accuracies for speakers in the bucket to be removed/re-registered are displayed in a different color and different markers. For visibility purposes, the total accuracy of the remaining buckets are displayed with a different color and different markers.}
    \label{acc_unreg_rereg_sup}
\end{figure}

Fig. \ref{acc_unreg_rereg_sup} shows the testing accuracy with respect to the required elapsed time for removal from and re-registration to a given bucket, e.g., in this case bucket $4$. The speaker(s) $[20]$, $[20, 21]$, and $[20, 21, 22]$ are efficiently removed from and re-registered to the bucket $4$ using the proposed method. As the the performance is measured on the testing utterances for all the $5$ speakers in the bucket, the testing accuracy drops by $\{20, 40, 60\}\%$ after removing one, two, and three speakers from the bucket, respectively. In other words, the proposed Algorithm. \ref{unreg_rereg_cnst_emb_rep} efficiently loads the already trained checkpoints for feature extraction of all the other buckets and the bucket for removing/re-registering speakers together with the corresponding checkpoints for the classifier. This leads to fast convergence and breaking the contrastive training for extraction of speaker equivariance inductive bias especially for the remaining buckets. In particular, removing and re-registering the aforementioned speaker(s) require approximately $\{1.71, 1.88, 1.91\}$ minutes for removing and $\{1.6, 2.7, 3.95\}$ minutes for re-registering the speaker(s) $[20]$, $[20, 21]$, and $[20, 21, 22]$, respectively, for the supervised case. For the case of unsupervised removal and re-registration, the elapsed times are approximately $\{1, 2, 2.5\}$ minutes for removing the speaker(s) $[20]$, $[20, 21]$, and $[20, 21, 22]$ and $\{2.5, 2.8, 3.3\}$ minutes for re-registering the speaker(s) $[20]$, $[20, 21, 22]$, and $[20, 21]$, respectively.

For the case of removing the entire speakers from the bucket, it is sufficient not to use the checkpoints of the trained contrastive feature encoder of that bucket, and proceed the training without providing the data from that bucket according to the Algorithm. \ref{unreg_rereg_cnst_emb_rep}. This results forgetting the contrastive inductive bias of the speakers in the bucket after approximately $2.4$ minutes, and consequently protecting them against re-identification. Re-registering the removed bucket takes approximately $2.6$ minutes. For the case of removing $4$ speakers from the bucket, it is possible to re-register the remaining speaker in another bucket if available or another agent with available bucket. This leads to forgetting the inductive bias of the $4$ speakers in the bucket by not providing  the data and the corresponding checkpoints of that bucket for those speakers during the training, and absorbing the remaining speaker in another bucket. Consequently, the problem is reduced to the case that all the speakers in the bucket are removed.

\begin{table}[tbp] 
\centering
\setlength{\tabcolsep}{3.5pt}     
\setlength{\cmidrulekern}{0.5em} 
\caption{Verification performance in terms of EER, $\mathrm{minDCF}$, and $\mathrm{minC}_{llr}$}
\begin{tabular}{%
  S[table-format=1.0]
  S[table-format=1.2]
  *{2}{ 
    *{2}{S[table-format=1.3]} 
    S[table-format=1.3]
  }
}
  \toprule
  & & \multicolumn{3}{c}{Supervised}     
    & \multicolumn{3}{c}{Unsupervised}    \\
    \cmidrule(lr){3-5}                  
    \cmidrule(lr){6-8}
  {\#} && {$\mathrm{EER}\:(\%)$} & {$\mathrm{minDCF}$} & {$\mathrm{minC}_{llr}$}  
                       & {$\mathrm{EER}\:(\%)$} & {$\mathrm{minDCF}$} & {$\mathrm{minC}_{llr}$} \\
  \midrule
  0 & & 0.983 & 0.075 & 0.041 & 1.068 & 0.118 & 0.053 \\
  1 & & 0.727 & 0.063 & 0.036 & 1.239 & 0.134 & 0.051 \\
  2 & & 0.855 & 0.095 & 0.037 & 1.239 & 0.122 & 0.05 \\
  3 & & 0.855 & 0.078 & 0.038 & 0.983 & 0.113 & 0.047 \\
  4 & & 1.154 & 0.117 & 0.05 & 1.45 & 0.175 & 0.071 \\
  \bottomrule
\end{tabular}
\label{verification_tab}
\end{table}

Table. \ref{verification_tab} reports the verification performance for different agents, i.e., Agents 0--4, in terms of \ac{EER} in $\%$, \ac{minDCF}, and minimum cost of log likelihood ratio calibration, $\mathrm{minC}_{llr}$, during the testing phase for the supervised and the unsupervised methods. The performance is reported after the completion of training procedure in Algorithm. \ref{cnst_emb_rep} for totally different hold-out utterances during the test time. To analyze the verification performance in the testing phase, the entire test samples are used. It is observed that the supervised mode always outperforms the unsupervised mode in terms of verification capabilities. This has to do with the additional information provided by the labels during the training. Moreover, the efficient use of the labelled data leads to a generally faster convergence and a better generalization capability during the inference and hence a better verification performance.

\section{Conclusions}\label{conclusions}
In this paper, an efficient method for consent management of speakers in the context of voice assistant systems is proposed. The proposed algorithms significantly reduce the convergence time of speaker recognition for consent management and outperform the baseline. Moreover, the proposed approach dynamically adapts to the consent status of each speaker. In other words, the process for registering new speakers, removing from the pool of registered speakers, and re-registering the speakers during the consent management are accomplished in a fast, dynamic, and memory efficient way. Furthermore, the proposed approach only requires a portion of utterances from the old registrations during new registrations leading to an improved privacy preservation. Finally, the proposed approach provides an improved verification performance in the supervised mode. 

\appendices
\section{Progressive Multi-Strided Random Buffer Sampling}\label{app:multi-stride-sampling}

\begin{algorithm}[!t]\label{alg_mult_str_updated}

\DontPrintSemicolon

\SetKwFunction{nutts}{$num_{spk,utts}$}
\SetKwFunction{ncollection}{$collection_{indx}$}
\SetKwFunction{interbkt}{$sample_{int-bkt}$}

\SetKwProg{fn}{def}{:}{}

\fn{\nutts{$max_{\mathrm{mem}}$, $\mathbf{n}_{bkt}$, $\mathbf{n}^{\mathrm{reg}}_{bkt}$}}{
$n_{\mathrm{tot}} = 0$\;
\For {$n_{b}$, $n^{\mathrm{reg}}_{b}$ in zip($\mathbf{n}_{bkt}$, $\mathbf{n}^{\mathrm{reg}}_{bkt}$)}{
$n_{\mathrm{tot}}$ $+=$ ($n_{b}$ $+$ $n^{\mathrm{reg}}_{b}$)
}

$n_{spk,\mathrm{utt}} = \lfloor max_{\mathrm{mem}} / n_{\mathrm{tot}} \rfloor $

\KwRet $n_{spk,\mathrm{utt}}$\;
}

\fn{\ncollection{$n_{s,\mathrm{utt}}$, $n_{spk,\mathrm{utt}}$, $\mathbf{n}_{bkt}$, $\mathbf{n}^{\mathrm{reg}}_{bkt}$}}{
$\mathcal{C}_{indx} = $ \{\}\;
\For {$n_{b}$, $n^{\mathrm{reg}}_{b}$ in zip($\mathbf{n}_{bkt}$, $\mathbf{n}^{\mathrm{reg}}_{bkt}$)}{
$\tilde{n}_{b} = n_{b} + n^{\mathrm{reg}}_{b}$\;
$\mathbf{indx}_{\mathrm{utt},b}$ $=$ $[\mathrm{rand}_{\mathrm{sample}}\left(\mathbf{indx}^{(i)}_{\mathrm{utt}}, n_{spk,\mathrm{utt}}\right)$ $\:\: for \:\: i \:\: in \:\: range(\tilde{n}_{b})]$\;

$\overline{\mathbf{indx}}_{\mathrm{utt},b} = \left[u \:\: for \:\: u_s \:\:  in \:\: \mathbf{indx}_{\mathrm{utt},b} \:\: for \:\: u \:\: in \:\: u_s\right]$

$\mathcal{C}_{indx}[b]\gets\overline{\mathbf{indx}}_{\mathrm{utt},b}$

}

\KwRet $\mathcal{C}_{indx}$\;
}

\fn{\interbkt{$\mathcal{C}_{indx}[b]$, $\mathbf{zy}_{b}$, $\mathbf{zy}_{\mathrm{init}}$, perm=True}}{

\If{perm}{
 $\mathcal{C}_{indx}[b] \gets \mathrm{rand}_{\mathrm{sample}}\left(\mathcal{C}_{indx}[b], len(\mathcal{C}_{indx}[b])\right)$
}

$\mathbf{z}_{\mathrm{init}}$, $\mathbf{y}_{\mathrm{init}}$ $=$ $\mathbf{zy}_{\mathrm{init}}$\;
$\mathbf{z}_{b}$, $\mathbf{y}_{b}$ $=$ $\mathbf{zy}_{b}$\;

$\mathbf{z}_{\mathrm{init}}.append(\mathbf{z}_{b}[\mathcal{C}_{indx}[b]])$\;
$\mathbf{y}_{\mathrm{init}}.append(\mathbf{y}_{b}[\mathcal{C}_{indx}[b]])$\;

$\mathbf{D}^{max_{\mathrm{mem}}}_{\mathrm{buff}} \gets concat(\mathbf{z}_{\mathrm{init}}, dim=0)$\;
$\mathbf{y}^{max_{\mathrm{mem}}}_{\mathrm{buff}} \gets concat(\mathbf{y}_{\mathrm{init}}, dim=0)$\;

\KwRet $\mathbf{D}^{max_{\mathrm{mem}}}_{\mathrm{buff}}$, $\mathbf{y}^{max_{\mathrm{mem}}}_{\mathrm{buff}}$\;
}

\caption{Functions for progressive multi-strided random buffer sampling}
\end{algorithm}

In Algorithm. \ref{alg_mult_str_updated}, $num_{spk,utts}(.)$ first computes the total number of speakers $n_{\mathrm{tot}}$ by looping through the zipped lists of the number of speakers per buckets $\mathbf{n}_{bkt}$ and the number of new speakers per buckets $\mathbf{n}^{\mathrm{reg}}_{bkt}$, in steps 3--5. Then, the number of utterances per speaker $n_{spk,\mathrm{utt}}$ is obtained by dividing the maximum allowed memory $max_{\mathrm{mem}}$ by the total number of speakers $n_{\mathrm{tot}}$, finding the floor of the division and converting the result to an integer using $\lfloor . \rfloor$ in step 6, and returning the result in step 7.

The function $collection_{indx}(.)$ loops through the zipped lists of the number of speakers per buckets $\mathbf{n}_{bkt}$ and the number of new speakers per buckets $\mathbf{n}^{\mathrm{reg}}_{bkt}$ in step 10. Then, the number of speakers per bucket is updated in step 11. Subsequently, the list comprehension of the indices of speakers' utterances in bucket $b$ is obtained in step 12. This is achieved using the updated number of speakers per bucket $\tilde{n}_{b}$ and the number of random selected utterances per speaker $n_{s,\mathrm{utt}}$ for the $i$-th speaker in the bucket defined as $\mathbf{indx}^{(i)}_{\mathrm{utt}} := [i.n_{s,\mathrm{utt}}, i.n_{s,\mathrm{utt}}+n_{s,\mathrm{utt}})$. The function $\mathrm{rand}_{\mathrm{sample}}(.)$ in step 12 randomly samples $n_{spk,\mathrm{utt}}$ utterances from $\mathbf{indx}^{(i)}_{\mathrm{utt}}$ with/without replacement. The resulting list comprehension for the indices of speaker(s) per bucket is flattened in step 13, and provided as the values of the dictionary $\mathcal{C}_{indx}$ for the given key $b$ in step 14. After the completion of the iterations for all the buckets, i.e., steps 10--15, the collection of indices of utterances for speaker(s) per bucket(s) is returned as a dictionary in step 16.

The function $sample_{int-bkt}(.)$ starts by random sampling of the collection of indices of utterances for a given bucket $b$ when the permutation $perm$ is set to True by default. This is shown in steps 18--20 where the operation $len(.)$ computes the length of a list. The initial tuple of speaker embeddings and corresponding labels is unpacked in step 21. Similarly, the tuple of speaker embeddings and corresponding labels for bucket $b$ is unpacked in step 22. Speaker embeddings and corresponding labels with the collection of indices of utterances for speakers in bucket $b$, shown as $\mathbf{z}_{b}[\mathcal{C}_{indx}[b]]$ and $\mathbf{y}_{b}[\mathcal{C}_{indx}[b]]$, respectively, are appended to the initial speaker embeddings $\mathbf{z}_{\mathrm{init}}$ and initial labels $\mathbf{y}_{\mathrm{init}}$ in steps 23 and $24$, respectively. The appended values from the previous steps are concatenated over the first dimension, i.e., the batch dimension of tensors or $dim=0$, for the progressive features $\mathbf{D}^{max_{\mathrm{mem}}}_{\mathrm{buff}}$ and the corresponding labels $\mathbf{y}^{max_{\mathrm{mem}}}_{\mathrm{buff}}$ with the maximum allowed memory size of $max_{\mathrm{mem}}$ in steps 25 and 26, respectively, and returned in step 27.

\section{Computing Optimal New Speakers/Buckets for Dynamic Registration}\label{app:opt-unique-bkts}

\begin{algorithm}[!t]\label{fn_dyn_cnst_emb_rep}

\DontPrintSemicolon

\SetKwFunction{uoptbkt}{$opt_{spk,bkt}$}

\SetKwProg{fn}{def}{:}{}

\fn{\uoptbkt{$\bar{\mathbf{z}}^{\mathrm{eval}}$, $\tilde{\mathbf{b}}$, $\mathbf{s}_{\mathrm{reg}}$, $\widetilde{\mathbf{s}}^{u}_{\mathrm{reg}}$, $n_{\mathrm{round}}$} }{

$\mathbf{z}^{\mathrm{eval}}$, $\mathbf{z}^{\mathrm{eval}}_{\mathrm{new}}$ = $\bar{\mathbf{z}}^{\mathrm{eval}}$

\If{$n_{\mathrm{round}}$ $==$ $0$}{
$\mathbf{b}^{*}_{\mathrm{sofar}}$, $\mathbf{s}^{*}_{\mathrm{sofar}}$ = [], []
}
\If{$n_{\mathrm{round}}$ $>$ $0$}{

Append $\widetilde{\mathbf{b}}^{*}_{\mathrm{reg}}$ from round $n_{\mathrm{round}}-1$ to $\mathbf{b}^{*}_{\mathrm{sofar}}$.\;
Append $\widetilde{\mathbf{s}}^{*}_{\mathrm{reg}}$ from round $n_{\mathrm{round}}-1$ to $\mathbf{s}^{*}_{\mathrm{sofar}}$.\;

}

\For{$s_{\mathrm{reg}}$ in $\mathbf{s}_{\mathrm{reg}}\backslash\widetilde{\mathbf{s}}^{u}_{\mathrm{reg}}$}{
\For {$\_$, $b$ in $enumerate(\tilde{\mathbf{b}})$}{

Compute \eqref{prototypes} for speakers in $b$ based on $\mathbf{z}^{\mathrm{eval}}_{b}$.

Compute \eqref{L2_distance} for $s_{\mathrm{reg}}$-th new speaker based on previous step and $\mathbf{z}^{\mathrm{eval}}_{\mathrm{new},\:s_{\mathrm{reg}},\:b}$.
}
Select the bucket according to \eqref{opt_bucket}.
}

Form $\mathbf{b}^{*}_{\mathrm{reg}}$.\;

Obtain $\widetilde{\mathbf{b}}^{*}_{\mathrm{reg}}$ and $\widetilde{\mathbf{s}}^{*}_{\mathrm{reg}}$ based on the Algorithm. \ref{dp_LUB_updated}.

\KwRet $\widetilde{\mathbf{b}}^{*}_{\mathrm{reg}}$, $\widetilde{\mathbf{s}}^{*}_{\mathrm{reg}}$, $\mathbf{b}^{*}_{\mathrm{sofar}}$, $\mathbf{s}^{*}_{\mathrm{sofar}}$
}

\caption{Compute optimal new speakers/buckets}
\end{algorithm}

In Algorithm. \ref{fn_dyn_cnst_emb_rep}, $opt_{spk,bkt}(.)$ provides a method to compute $\widetilde{\mathbf{b}}^{*}_{\mathrm{reg}}$, $\widetilde{\mathbf{s}}^{*}_{\mathrm{reg}}$, $\mathbf{b}^{*}_{\mathrm{sofar}}$, and $\mathbf{s}^{*}_{\mathrm{sofar}}$. First the tuple of speaker embeddings for evaluation is unpacked to obtain the evaluation embeddings according to the old and the new datasets in step 2. In case the round number is zero, $n_{\mathrm{round}} = 0$, $\mathbf{b}^{*}_{\mathrm{sofar}}$ and $\mathbf{s}^{*}_{\mathrm{sofar}}$ are set to empty lists in step 4. Otherwise, for the round number $n_{\mathrm{round}} > 0$, $\mathbf{b}^{*}_{\mathrm{sofar}}$ and $\mathbf{s}^{*}_{\mathrm{sofar}}$ are appended by $\widetilde{\mathbf{b}}^{*}_{\mathrm{reg}}$ and $\widetilde{\mathbf{s}}^{*}_{\mathrm{reg}}$ from the round $n_{\mathrm{round}} - 1$ in steps 7 and 8, respectively. For each new speaker $s_{\mathrm{reg}}$ in the updated set of new speakers excluding the already registered new speakers, $\mathbf{s}_{\mathrm{reg}}\backslash\widetilde{\mathbf{s}}^{u}_{\mathrm{reg}}$, step 10, and all the buckets $\tilde{\mathbf{b}}$, step 11, compute the prototypes in the inference mode, step 12, as follows.
\begin{equation}\label{prototypes}
\mathbf{c}_{s,b}=\frac{1}{\lvert \widetilde{\mathcal{P}}_{s,b} \rvert}\sum_{i\in\widetilde{\mathcal{P}}_{s,b}}\mathbf{z}^{(i)}_{s,b},
\end{equation}
where $\mathbf{c}_{s,b}$ denotes the prototype of speaker $s$ in bucket $b$, $\widetilde{\mathcal{P}}_{s,b}$ denotes the set of hold-out utterances during the inference for the speaker $s$ in bucket $b$ with the cardinality of $\lvert \widetilde{\mathcal{P}}_{s,b} \rvert$, and $\mathbf{z}^{(i)}_{s,b}$ is the corresponding embedding following the same definition as in \eqref{enc}. Subsequently, the $L2$ pairwise distance of the encoded features of $n_{\mathrm{reg}}$-th new speaker in the bucket $b$, $\mathbf{z}_{s_{n_{\mathrm{reg}}},b}$, from the prototypes $\mathbf{c}_{s,b}$ is calculated for the hold-out utterances of the new speaker for $\forall n_{\mathrm{reg}}$, $\forall b$, and $\forall s$ in step 13 as follows.
\begin{equation}\label{L2_distance}
d(\mathbf{z}_{s_{n_{\mathrm{reg}}},b}, \mathbf{c}_{s,b})= \|\mathbf{z}_{s_{n_{\mathrm{reg}}},b}-\mathbf{c}_{s,b}\|^{2}, 
\end{equation}
where $d(\mathbf{z}_{s_{n_{\mathrm{reg}}},b}, \mathbf{c}_{s,b})$ denotes the Euclidean distance. Consequently, the corresponding label of the bucket including the speaker's prototype with the shortest $L2$ pairwise distance from the new registered speaker is returned for $\forall n_{\mathrm{reg}}$ in step 15 as follows.
\begin{equation}\label{opt_bucket}
\_,b^{*}_{n_{\mathrm{reg}}}=\operatorname*{argmin}_{s,b} d(\bar{\mathbf{z}}_{s_{n_{\mathrm{reg}}},b}, \mathbf{c}_{s,b}),
\end{equation}
where $b^{*}_{n_{\mathrm{reg}}}$ represents the optimal bucket index for the registration of $n_{\mathrm{reg}}$-th new speaker. The term $\bar{\mathbf{z}}_{s_{n_{\mathrm{reg}}},b}$ in \eqref{opt_bucket} denotes the embedding of the $n_{\mathrm{reg}}$-th new registered speaker $s_{n_{\mathrm{reg}}}$ in bucket $b$ averaged over the corresponding utterances as follows.
\begin{equation}\label{mean_new_spk}
\bar{\mathbf{z}}_{s_{n_{\mathrm{reg}}},b}=\frac{1}{\lvert \widetilde{\mathcal{P}}_{s_{n_{\mathrm{reg}}},b} \rvert}\sum_{i\in\widetilde{\mathcal{P}}_{s_{n_{\mathrm{reg}}},b}}\mathbf{z}^{(i)}_{s_{n_{\mathrm{reg}}},b},
\end{equation}
in which $\widetilde{\mathcal{P}}_{s_{n_{\mathrm{reg}}},b}$ denotes the set of hold-out utterances during the inference for the new speaker $s_{n_{\mathrm{reg}}}$. Subsequently, the index of optimal buckets for all new speaker registrations in the current round obtained according to \eqref{prototypes}-\eqref{opt_bucket} forms $\mathbf{b}^{*}_{\mathrm{reg}}$ in step 17. Consequently, a dynamic programming approach of decision type is designed in Algorithm. \ref{dp_LUB_updated} to make sure in each round a subset of unique new speakers $\tilde{\mathbf{s}}^{*}_{\mathrm{reg}}$ are registered in the sequence of longest optimal unique buckets $\widetilde{\mathbf{b}}^{*}_{\mathrm{reg}}$ and do not share the same bucket, step 18 in Algorithm. \ref{fn_dyn_cnst_emb_rep}.

\begin{algorithm}[!t]\label{dp_LUB_updated}

\SetAlgoLined
\DontPrintSemicolon

\nonl\textbf{Input:} The sequence of optimal buckets for new speaker registrations $\mathbf{b}^{*}_{\mathrm{reg}}$.

\nonl\textbf{Output:} The sequence of longest optimal unique per registration buckets $\tilde{\mathbf{b}}^{*}_{\mathrm{reg}}$ and the corresponding index of new registered speakers $\tilde{\mathbf{s}}^{*}_{\mathrm{reg}}$.

\nonl\textbf{Subproblem:} The sequence of longest optimal unique per registration buckets $\tilde{\mathbf{b}}^{*}_{\mathrm{reg}}[:i]$ for the set of new speakers in the interval $[N,i)$ for $i\in\{N,\ldots,N+len(\mathbf{s}_{\mathrm{reg}})-1\}$.

\nonl\textbf{Relation:} Recursive computations to obtain the sequence of longest optimal unique per registration buckets in \eqref{relate}.

\nonl\textbf{Topological} \textbf{Order:} Sub-problem $\tilde{\mathbf{b}}^{*}_{\mathrm{reg}}[:i+1]$ only depends on strictly smaller $i$, so it is acyclic, i.e., increase $i$ for $i=N,\ldots,N+len(\mathbf{s}_{\mathrm{reg}})-1$.

\nonl\textbf{Base} \textbf{Case:} The empty set is always achieved for $\tilde{\mathbf{b}}^{*}_{\mathrm{reg}}[:N]=\emptyset$.

\nonl\textbf{Original} \textbf{Problem:} The sequence of longest optimal unique per registration buckets for the entire set of new speakers, i.e., $\tilde{\mathbf{b}}^{*}_{\mathrm{reg}}[:N+len(\mathbf{s}_{\mathrm{reg}})]$.

\caption{Longest unique sequence of optimal buckets per registration}
\end{algorithm}

The dynamic programming Algorithm. \ref{dp_LUB_updated} of decision type obtains the longest unique sequence of the optimal buckets for registering new speakers in each round. It receives the full list of optimal buckets for the current round $\mathbf{b}^{*}_{\mathrm{reg}}$ as the input. Then, the longest unique sequence of the optimal buckets is achieved according to the recursive call as follows. 
\begin{equation}\label{relate}
\widetilde{\mathbf{b}}^{*}_{\mathrm{reg}}[:i+1] = \left\{ \,
\begin{IEEEeqnarraybox}[][c]{l?s}
\IEEEstrut
\widetilde{\mathbf{b}}^{*}_{\mathrm{reg}}[:i] & $b^{*}_{i}\in\widetilde{\mathbf{b}}^{*}_{\mathrm{reg}}[:i]$,
\\
\widetilde{\mathbf{b}}^{*}_{\mathrm{reg}}[:i]\cup\{b^{*}_{i}\} & $b^{*}_{i}\notin\widetilde{\mathbf{b}}^{*}_{\mathrm{reg}}[:i]$.
\IEEEstrut
\end{IEEEeqnarraybox}
\right.
\end{equation}
In \eqref{relate}, the optimal bucket at index $i$, $b^{*}_{i}$, is added only if it does not already exist in the set of new speakers in the interval $[0,i)$. The standard steps for the dynamic process to find the solution for the subset of the original problem using the subproblem for the base case and the relation in \eqref{relate} are described in the Algorithm. \ref{dp_LUB_updated}. Consequently, by increasing the index $i$ the entire list of optimal buckets is covered starting from the base case in the bottom-up way. The proposed dynamic programming algorithm only requires the linear time complexity of $O(N_{\mathrm{reg}})$ for the worst case, i.e., $len(\mathbf{s}_{\mathrm{reg}})=N_{\mathrm{reg}}$. This is due to the fact that the sequence of new speaker registrations is progressively reduced after each round. 


\section{Computing Properties for Registering}\label{app:prop-reg}

\begin{algorithm}[!t]\label{fn_dyn_spk_bkt_pattern}

\DontPrintSemicolon
\SetKwFunction{ptrna}{$\mathrm{strategy}_{1}$}
\SetKwFunction{ptrnb}{$\mathrm{strategy}_{2}$}
\SetKwFunction{ptrnc}{$\mathrm{strategy}_{3}$}
\SetKwFunction{ptrnd}{$\mathrm{strategy}_{4}$}

\SetKwFunction{strategyslctreg}{$\mathrm{strategy}^{reg}_{selct}$}

\SetKwFunction{perroundsspksbkts}{$prop^{reg}_{spk,bkt}$}

\SetKwProg{fn}{def}{:}{}

\fn{\ptrna{$\widetilde{\mathbf{s}}^{*}_{\mathrm{reg}}$, $\widetilde{\mathbf{b}}^{*}_{\mathrm{reg}}$, $b$, $n_{b}$}}{

\KwRet $\widetilde{\mathbf{s}}^{*}_{\mathrm{reg}}[find(b\in\widetilde{\mathbf{b}}^{*}_{\mathrm{reg}})], \mathrm{pattern}_{1}$,
$n_{b}$, $1$
}
\fn{\ptrnb{$\mathbf{s}^{*}_{\mathrm{sofar}}$, $\mathbf{b}^{*}_{\mathrm{sofar}}$, $b$, $n_{b}$}}{
\KwRet $(\mathbf{s}^{*}_{\mathrm{sofar}}[find(b\in\mathbf{b}^{*}_{\mathrm{sofar}})]$, $\mathrm{pattern}_{2}$, $n_{b}+len(\mathbf{s}^{*}_{\mathrm{sofar}}[find(b\in\mathbf{b}^{*}_{\mathrm{sofar}})])$, $0)$
}
\fn{\ptrnc{$\widetilde{\mathbf{s}}^{*}_{\mathrm{reg}}$, $\widetilde{\mathbf{b}}^{*}_{\mathrm{reg}}$, $\mathbf{s}^{*}_{\mathrm{sofar}}$, $\mathbf{b}^{*}_{\mathrm{sofar}}$, $b$, $n_{b}$}}{
\KwRet $(\widetilde{\mathbf{s}}^{*}_{\mathrm{reg}}[find(b\in\widetilde{\mathbf{b}}^{*}_{\mathrm{reg}})]\cup\mathbf{s}^{*}_{\mathrm{sofar}}[find(b\in\mathbf{b}^{*}_{\mathrm{sofar}})]$, $\mathrm{pattern}_{3}$,
$n_{b}+len(\mathbf{s}^{*}_{\mathrm{sofar}}[find(b\in\mathbf{b}^{*}_{\mathrm{sofar}})])$, $1)$
}
\fn{\ptrnd{$n_{b}$}}{
\KwRet [], $\mathrm{pattern}_{4}$, $n_{b}$, $0$
}

\fn{\strategyslctreg{$b$, $\widetilde{\mathbf{b}}^{*}_{\mathrm{reg}}$, $\mathbf{b}^{*}_{\mathrm{sofar}}$, $\widetilde{\mathbf{s}}^{*}_{\mathrm{reg}}$, $\mathbf{s}^{*}_{\mathrm{sofar}}$}}{

	$\mathrm{pattern}_{1} = partial(\mathrm{strategy}_{1}, \widetilde{\mathbf{s}}^{*}_{\mathrm{reg}}, \widetilde{\mathbf{b}}^{*}_{\mathrm{reg}}, b)$\;
	$\mathrm{pattern}_{2} = partial(\mathrm{strategy}_{2}, \mathbf{s}^{*}_{\mathrm{sofar}}, \mathbf{b}^{*}_{\mathrm{sofar}}, b)$\;
	$\mathrm{pattern}_{3} = partial(\mathrm{strategy}_{3}, \widetilde{\mathbf{s}}^{*}_{\mathrm{reg}}, \widetilde{\mathbf{b}}^{*}_{\mathrm{reg}}, \mathbf{s}^{*}_{\mathrm{sofar}}, \mathbf{b}^{*}_{\mathrm{sofar}}, b)$\;
	$\mathrm{pattern}_{4} = \mathrm{strategy}_{4}$

	pattern $=$ \{$\mathrm{pattern}_{1}$: $( b\in\widetilde{\mathbf{b}}^{*}_{\mathrm{reg}}$ \& $b\notin\mathbf{b}^{*}_{\mathrm{sofar}})$,
	$\mathrm{pattern}_{2}$: $( b\notin\widetilde{\mathbf{b}}^{*}_{\mathrm{reg}}$ \& $b\in\mathbf{b}^{*}_{\mathrm{sofar}})$,
	$\mathrm{pattern}_{3}$: $( b\in\widetilde{\mathbf{b}}^{*}_{\mathrm{reg}}$ \& $b\in\mathbf{b}^{*}_{\mathrm{sofar}})$,
	$\mathrm{pattern}_{4}$: $( b\notin\widetilde{\mathbf{b}}^{*}_{\mathrm{reg}}$ \& $b\notin\mathbf{b}^{*}_{\mathrm{sofar}})$\}
	
\For{$\mathrm{pattern}_{selct}$, $\mathrm{logic}$ in $\mathrm{pattern}$.items()}{
\If{$\mathrm{logic}$}{
\KwRet $\mathrm{pattern}_{selct}$
}
}
}

\fn{\perroundsspksbkts{$\tilde{\mathbf{b}}$, $\mathbf{n}_{bkt}$, $\widetilde{\mathbf{b}}^{*}_{\mathrm{reg}}$, $\mathbf{b}^{*}_{\mathrm{sofar}}$, $\widetilde{\mathbf{s}}^{*}_{\mathrm{reg}}$, $\mathbf{s}^{*}_{\mathrm{sofar}}$}}{
$\tilde{\mathbf{n}}_{bkt}$, $\tilde{\mathbf{n}}^{\mathrm{reg}}_{bkt}$, $\mathcal{S}_{\mathrm{reg}}$, $\mathcal{P}_{\mathrm{reg}}$ $=$ [], [], \{\}, \{\}\;
\For{$b$, $n_{b}$ in $zip(\tilde{\mathbf{b}},\:\mathbf{n}_{bkt})$}{
$\mathrm{pattern}_{selct}$ $=$ \strategyslctreg{$b$, $\widetilde{\mathbf{b}}^{*}_{\mathrm{reg}}$, $\mathbf{b}^{*}_{\mathrm{sofar}}$, $\widetilde{\mathbf{s}}^{*}_{\mathrm{reg}}$, $\mathbf{s}^{*}_{\mathrm{sofar}}$}\;
$\mathcal{S}_{\mathrm{reg}}[b]$, $\mathcal{P}_{\mathrm{reg}}[b]$, $\tilde{\mathbf{n}}_{bkt}[b]$, $\tilde{\mathbf{n}}^{\mathrm{reg}}_{bkt}[b]$ $\gets$ $\mathrm{pattern}_{selct}$($n_{b}$)
}

\KwRet $\tilde{\mathbf{n}}_{bkt}$, $\tilde{\mathbf{n}}^{\mathrm{reg}}_{bkt}$, $\mathcal{S}_{\mathrm{reg}}$, $\mathcal{P}_{\mathrm{reg}}$\;

}

\caption{Compute properties for registering}
\end{algorithm}

In Algorithm. \ref{fn_dyn_spk_bkt_pattern}, $prop^{reg}_{spk,bkt}(.)$  provides the required properties for dynamic registrations of the new speakers. First, the required properties including $\tilde{\mathbf{n}}_{bkt}$, $\tilde{\mathbf{n}}^{\mathrm{reg}}_{bkt}$, $\mathcal{S}_{\mathrm{reg}}$, and $\mathcal{P}_{\mathrm{reg}}$ are initialized by empty lists and empty dictionaries, respectively, in step 21. Then, the function loops through the zipped lists of buckets $\tilde{\mathbf{b}}$ and corresponding number of speakers in buckets $\mathbf{n}_{bkt}$ in step 22. Subsequently, for each bucket $b$ there exists four different patterns/strategies provided by the function $\mathrm{strategy}^{reg}_{selct}(.)$ defined in step 9. The function $\mathrm{strategy}^{reg}_{selct}(.)$ first unifies the specific arguments for different strategies through the $partial(.)$ operation, similar to the $partial(.)$ in Python, shown in steps 10--13. The aforementioned patterns are used as the keys for pattern dictionary with the values representing different logics for registration in step 14. Looping through the dictionary of the pattern from the previous step, the appropriate pattern is selected and returned if the corresponding logic is fulfilled, steps 15--19.

The specific functions for different strategies are defined in steps 1--8. The first strategy is selected, if $b$ belongs to $\widetilde{\mathbf{b}}^{*}_{\mathrm{reg}}$ and not to $\mathbf{b}^{*}_{\mathrm{sofar}}$. The set of indices of new speakers in which $b\in\widetilde{\mathbf{b}}^{*}_{\mathrm{reg}}$ is obtained using $find(.)$ operation as the first term to return in step 2. The corresponding pattern status of $\mathrm{pattern}_{1}$ is returned as the second term, the number of speakers per bucket in this case $n_{b}$ is  returned as the third term, and finally the number of new speaker to be registered in this bucket under this strategy is one. Similarly, the rest of strategies are selected based on the corresponding logic, defined in step 14, and the desired properties are returned, steps 4, 6, and 8.

\section{Computing Properties for Removing}\label{app:prop-unreg}

\begin{algorithm}[t]\label{fn_unreg_spk_bkt_pattern}

\DontPrintSemicolon

\SetKwFunction{ptrna}{$\mathrm{strategy}_{1}$}
\SetKwFunction{ptrnb}{$\mathrm{strategy}_{2}$}

\SetKwFunction{strategyslctunreg}{$\mathrm{strategy}^{unreg}_{selct}$}

\SetKwFunction{unregspksbkts}{$prop^{unreg}_{spk,bkt}$}

\SetKwProg{fn}{def}{:}{}

\fn{\ptrna{$b$, $\tilde{\mathbf{b}}_{\mathrm{unreg}}$, $\mathbf{s}_{\mathrm{res}}$}}{

\KwRet $\mathbf{s}_{\mathrm{res}}[find(b\in\tilde{\mathbf{b}}_{\mathrm{unreg}})]$, $\mathrm{pattern}_{1}$, $len(\mathbf{s}_{\mathrm{res}}[find(b\in\tilde{\mathbf{b}}_{\mathrm{unreg}})])$
}
\fn{\ptrnb{$n_{b}$, $b$, $\tilde{\mathbf{b}}_{\mathrm{unreg}}$, $\mathbf{s}_{\mathrm{res}}$}}{
\KwRet $\mathbf{s}_{\mathrm{res}}[find(b\notin\tilde{\mathbf{b}}_{\mathrm{unreg}})]$, $\mathrm{pattern}_{2}$, $n_{b}$
}

\fn{\strategyslctunreg{$b$, $n_{b}$, $\tilde{\mathbf{b}}_{\mathrm{unreg}}$}}{

	$\mathrm{pattern}_{1} = \mathrm{strategy}_{1}$\;
	$\mathrm{pattern}_{2} = partial(\mathrm{strategy}_{2}, n_{b})$
	
	pattern $=$ \{$\mathrm{pattern}_{1}$: $b\in\tilde{\mathbf{b}}_{\mathrm{unreg}}$,
	$\mathrm{pattern}_{2}$: $b\notin\tilde{\mathbf{b}}_{\mathrm{unreg}}$\}
	
\For{$\mathrm{pattern}_{selct}$, $\mathrm{logic}$ in $\mathrm{pattern}$.items()}{
\If{$\mathrm{logic}$}{
\KwRet $\mathrm{pattern}_{selct}$
}
}
}

\fn{\unregspksbkts{$\tilde{\mathbf{b}}$, $\mathbf{n}_{bkt}$, $\tilde{\mathbf{b}}_{\mathrm{unreg}}$, $\mathbf{s}_{\mathrm{res}}$}}{
$\tilde{\mathbf{n}}_{bkt}$, $\mathcal{S}_{\mathrm{res}}$, $\mathcal{P}_{\mathrm{unreg}}$ $=$ [], \{\}, \{\}\;
\For{$b$, $n_{b}$ in $zip(\tilde{\mathbf{b}},\:\mathbf{n}_{bkt})$}{
$\mathrm{pattern}_{selct}$ $=$ \strategyslctunreg{$b$, $n_{b}$, $\tilde{\mathbf{b}}_{\mathrm{unreg}}$}\;
$\mathcal{S}_{\mathrm{res}}[b]$, $\mathcal{P}_{\mathrm{unreg}}[b]$, $\tilde{\mathbf{n}}_{bkt}[b]$ $\gets$ $\mathrm{pattern}_{selct}$($b$, $\tilde{\mathbf{b}}_{\mathrm{unreg}}$, $\mathbf{s}_{\mathrm{res}}$)
}

\KwRet $\tilde{\mathbf{n}}_{bkt}$, $\mathcal{S}_{\mathrm{res}}$, $\mathcal{P}_{\mathrm{unreg}}$\;

}

\caption{Compute properties for removing}
\end{algorithm}

In Algorithm. \ref{fn_unreg_spk_bkt_pattern}, $prop^{unreg}_{spk,bkt}(.)$ provides the required properties for removing the given set of speakers from the pool of already registered speakers. First, the required properties $\tilde{\mathbf{n}}_{bkt}$, $\mathcal{S}_{\mathrm{res}}$, and $\mathcal{P}_{\mathrm{unreg}}$ are initialized by an empty list and empty dictionaries, respectively, in step 15. Then, the function loops through the zipped lists of buckets $\tilde{\mathbf{b}}$ and corresponding number of speakers in buckets $\mathbf{n}_{bkt}$ in step 16. Subsequently, for each bucket $b$ there exists two different patterns/strategies provided by the function $\mathrm{strategy}^{unreg}_{selct}(.)$ defined in step 5. The function $\mathrm{strategy}^{unreg}_{selct}(.)$ first unifies the specific arguments for different strategies through the $partial(.)$ operation, shown in steps 6 and 7. The aforementioned patterns are used as the keys for pattern dictionary with the values representing different logics for removal in step 8. Looping through the dictionary of the pattern from the previous step, the appropriate pattern is selected and returned if the corresponding logic is fulfilled, steps 9--13.

The specific functions for different strategies are defined in steps 1--4. The first strategy is selected, if $b$ belongs to $\tilde{\mathbf{b}}_{\mathrm{unreg}}$. The set of indices of residual speakers in which $b\in\tilde{\mathbf{b}}_{\mathrm{unreg}}$ is obtained using $find(.)$ operation as the first term to return in step 2. The corresponding pattern status of $\mathrm{pattern}_{1}$ is returned as the second term, and the number of remaining speakers per bucket $\tilde{\mathbf{n}}_{bkt}[b]$, in this case $len(\mathbf{s}_{\mathrm{res}}[find(b\in\tilde{\mathbf{b}}_{\mathrm{unreg}})])$, is returned as the last term. Similarly, the desired properties for removing are returned if the condition for the first strategy is not fulfilled as shown in step 4. In this case, $\tilde{\mathbf{n}}_{bkt}[b]$ is set to the initial state before removal, i.e., $n_{b}$.

\section*{Acknowledgment}
This publication has emanated from research conducted with the financial support of Science Foundation Ireland under Grant number 19/FFP/6775.

\bibliographystyle{IEEEtran}
\bibliography{references}

\end{document}